%% file: main.tex
\documentclass[submission,copyright,creativecommons]{eptcs}
\providecommand{\event}{SNR 2021} % Name of the event you are submitting to

\usepackage{amsfonts}
\usepackage{graphicx}
\usepackage{enumerate}
\RequirePackage{graphics}
\usepackage{xcolor}
\usepackage{footnote}
\usepackage{xparse}
\usepackage{cite}
\usepackage{rotating}

\usepackage{amssymb}
\usepackage{amsmath}
\usepackage{keyval}
\usepackage{xspace}
\usepackage{paralist}
\usepackage{listings}
\usepackage{multirow}
\usepackage{stmaryrd}
\usepackage{adjustbox} % scaling figure size
\usepackage{makecell}

\RequirePackage{tikz}
\usetikzlibrary{arrows,automata,shapes,calc,through,decorations.pathmorphing,decorations.fractals,chains,shapes.multipart}

\usepackage{arydshln} % hdashline
\usepackage[free-standing-units]{siunitx}

\usepackage{circuitikz}

\include{prelude}

\title{Model Checking for Rectangular Hybrid Systems: A Quantified Encoding Approach}
\author{Luan V. Nguyen
\institute{Department of Computer Science\\ University of Dayton\\ OH, USA}
\email{lnguyen1@udayton.edu}
\and
Wesam Haddad
\institute{Department of Computer Science\\ University of Dayton\\ OH, USA}
\email{haddadw1@udayton.edu}
\and
Taylor T. Johnson
\institute{Department of Computer Science\\ Vanderbilt University\\ TN, USA}
\email{taylor.johnson@vanderbilt.edu}
}
\def\titlerunning{QBMC for RHA}
\def\authorrunning{Nguyen et al.}
\begin{document}
\maketitle

% correct bad hyphenation here
%\hyphenation{op-tical net-works semi-conduc-tor}


%\title{\titlename}


%\author{\authorluan,~\authortran,~and~\authortaylor \\ University of Texas at Arlington}%
%\author{\authorluan \inst{1} \and \authortaylor \inst{2}}

%\institute{Department of Computer Science, University of Texas at Arlington, USA\and University of Toronto, Canada \and Vanderbilt University, USA}
%\institute{Department of Computer Science and Engineering, University of Texas at Arlington, TX, USA \\ \email{luanvnguyen@mavs.uta.edu} \and Institute for Software Integrated Systems, Vanderbilt University, TN, USA \\ \email{taylor.johnson@vanderbilt.edu}}

%\institute{University of Dayton, Arlington, TX, USA \\ \email{luanvnguyen@mavs.uta.edu} \and Vanderbilt University, Nashville, TN, USA \\ \email{taylor.johnson@vanderbilt.edu}}

%\maketitle

\begin{abstract}
	\input{abstract}
\end{abstract}

\input{intro}

\input{rw}

\input{preliminary}
\input{algorithm}
%\vspace{1em}
\input{experiment}
\input{conclusion}
\vspace{-1em}
\section*{Acknowledgements}
\vspace{-1em}
The material presented in this paper is based upon work supported by the National Science Foundation (NSF) through grant numbers 1918450 and 2028001. Any opinions, findings, and conclusions or recommendations expressed in this paper are those of the authors and do not necessarily reflect the views of NSF. This paper is based on an earlier workshop paper that did not appear in an archival proceedings: Luan Viet Nguyen, Djordje Maksimovic, Taylor T. Johnson, Andreas Veneris, "Quantified Bounded Model Checking for Rectangular Hybrid Automata," which was presented at the 9th International Workshop on Constraints in Formal Verification (CFV 2015), Austin, Texas, November 2015, colocated with the 34th IEEE/ACM International Conference on Computer-Aided Design (ICCAD 2015).

% please use the full citation for it, even though no doi, etc: 
% https://web.engr.oregonstate.edu/~alex/cfv15.html

%\balance % equalize columns

%\scriptsize
%\normalsize
%\let\oldbibliography\thebibliography
%\renewcommand{\thebibliography}[1]{\oldbibliography{#1}
%\setlength{\itemsep}{0pt}} %Reducing spacing in the bibliography.
\bibliographystyle{eptcs}
\bibliography{master,luan}
%\bibliographystyle{splncs03}
%\bibliography{master,luan}  % sigproc.bib is the name of the Bibliography in this case
%\normalsize

%\clearpage

%\input{appendix}



%\newpage



%\vfill


\end{document}

%% file: prelude.tex
\newcommand{\commenttaylor}[1]{}

\newcommand{\nnnum}[1]{\relax\ifmmode 
  {\mathbb #1}_{\geq 0} \else ${\mathbb #1}_{\geq 0}$
  \fi}
\newcommand{\npnum}[1]{\relax\ifmmode 
  {\mathbb #1}_{\leq 0} \else ${\mathbb #1}_{\leq 0}$
  \fi}
\newcommand{\pnum}[1]{\relax\ifmmode 
  {\mathbb #1}_{> 0} \else ${\mathbb #1}_{> 0}$
  \fi}
\newcommand{\nnum}[1]{\relax\ifmmode 
  {\mathbb #1}_{< 0} \else ${\mathbb #1}_{< 0}$
  \fi}
\newcommand{\plnum}[1]{\relax\ifmmode 
  {\mathbb #1}_{+} \else ${\mathbb #1}_{+}$
  \fi}
\newcommand{\nenum}[1]{\relax\ifmmode 
  {\mathbb #1}_{-} \else ${\mathbb #1}_{-}$
  \fi}

\newcommand{\extb}[1]{\relax\ifmmode {\sf ExtBeh}_{#1} \else ${\sf ExtBeh}_{#1}$\fi} 
\newcommand{\tdists}[1]{\relax\ifmmode {\sf Tdists}_{#1} \else ${\sf Tdists}_{#1}$\fi} 

\newcommand{\exec}[1]{\relax\ifmmode {\sf Execs}_{#1} \else ${\sf Exec}_{#1}$\fi} 
\newcommand{\execf}[1]{\relax\ifmmode {\sf Execs}^*_{#1} \else ${\sf Exec}^*_{#1}$\fi} 
\newcommand{\execi}[1]{\relax\ifmmode {\sf Execs}^\omega_{#1} \else ${\sf Exec}^\omega_{#1}$\fi} 

\newcommand{\ctrace}[1]{\relax\ifmmode {\sf Ctraces}_{#1} \else ${\sf Ctraces}_{#1}$\fi} 

\newcommand{\trace}[1]{\relax\ifmmode {\sf Traces}_{#1} \else ${\sf Traces}_{#1}$\fi} 
\newcommand{\tracef}[1]{\relax\ifmmode {\sf Traces}^*_{#1} \else ${\sf Traces}^*_{#1}$\fi} 
\newcommand{\tracei}[1]{\relax\ifmmode {\sf Traces}^\omega_{#1} \else ${\sf Traces}^\omega_{#1}$\fi} 

\newcommand{\frag}[1]{\relax\ifmmode {\sf Frags}_{#1} \else ${\sf Frags}_{#1}$\fi} 
\newcommand{\fragf}[1]{\relax\ifmmode {\sf Frags}^*_{#1} \else ${\sf Frags}^*_{#1}$\fi} 
\newcommand{\fragi}[1]{\relax\ifmmode {\sf Frags}^\omega_{#1} \else ${\sf Frags}^\omega_{#1}$\fi} 

\newcommand{\reach}[1]{\relax\ifmmode {\sf Reach}_{#1} \else ${\sf Reach}_{#1}$\fi}

\def\A{{\cal A}} % HA
\def\E{{\cal E}} % HA
\def\H{{\cal H}} % HA
\def\I{{\cal I}} % environment sequence
\def\Q{{\cal Q}} % set of modes
\def\R{{\cal R}} % relation
\def\T{{\cal T}} % set of trajectories
\def\U{{\cal U}} % set of trajectories
\newcommand{\col}[1]{\relax\ifmmode \mathscr #1\else $\mathscr #1$\fi}

\definecolor{HIOAcolor}{rgb}{0.776,0.22,0.07}

\newcommand{\SC}[2]{\relax\ifmmode {\tt Scount}(#1,#2) \else ${\tt Scount}(#1,#2)$\fi} 
\newcommand{\SCM}[2]{\relax\ifmmode {\tt Smin}(#1,#2) \else ${\tt Smin}(#1,#2)$\fi} 
\newcommand{\Aut}[1]{\relax\ifmmode {\tt Aut}(#1) \else ${\tt Aut}(#1)$\fi}

\newcommand{\act}[1]{{\operatorname{\mathsf{#1}}}}

\newcommand{\deq}{\mathrel{\stackrel{\scriptscriptstyle\Delta}{=}}}
\newcommand{\seclabel}[1]{\label{sec:#1}}
\newcommand{\secref}[1]{Section~\ref{sec:#1}}

\newcommand{\figlabel}[1]{\label{fig:#1}}
\newcommand{\figref}[1]{Figure~\ref{fig:#1}}

\newcommand{\tablabel}[1]{\label{tab:#1}}
\newcommand{\tabref}[1]{Table~\ref{tab:#1}}

\newcommand{\eqlabel}[1]{\label{eq:#1}}
\renewcommand{\eqref}[1]{Equation~\ref{eq:#1}}

\newcommand{\remove}[1]{}
\newcommand{\salg}[1]{\relax\ifmmode {\mathcal F}_{#1}\else ${\mathcal F}_{#1}$\fi} 
\newcommand{\msp}[1]{\relax\ifmmode (#1, \salg{#1}) \else $(#1, \salg{#1})$\fi} 
\newcommand{\msprod}[2]{\relax\ifmmode ( #1 \times #2, \salg{#1} \otimes \salg{#2}) \else $(#1 \times #2, \salg{#1} \otimes \salg{#2})$\fi} 
\newcommand{\dist}[1]{\relax\ifmmode {\mathcal P}\msp{#1}
  \else ${\mathcal P}\msp{#1}$\fi} 
\newcommand{\subdist}[1]{\relax\ifmmode {\mathcal S}{\mathcal P}\msp{#1} 
  \else ${\mathcal S}{\mathcal P}\msp{#1}$\fi} 
\newcommand{\disc}[1]{\relax\ifmmode {\sf Disc}(#1)
  \else ${\sf Disc}(#1)$\fi} 

\newcommand{\Trajeq}{\relax\ifmmode {\mathcal R}_\T \else ${\mathcal R}_\T$\fi} 
\newcommand{\Acteq}{\relax\ifmmode {\mathcal R}_A \else ${\mathcal R}_A$\fi} 
\newcommand{\noop}{\relax\ifmmode \lambda \else $\lambda$\fi} 
\newcommand{\close}[1]{\relax\ifmmode \overline{#1} \else $\overline{#1}$\fi}

\newcommand{\ie}{i.e.,\xspace}

\newcommand{\eg}{e.g.,\xspace}

\newcommand{\tup}[1]
           {
             \relax\ifmmode
             \langle #1 \rangle
             \else $\langle$ #1 $\rangle$ \fi
           }

\newcommand{\lit}[1]{ \relax\ifmmode
                \mathord{\mathcode`\-="702D\sf #1\mathcode`\-="2200}
                \else {\it #1} \fi }

\newcommand{\figuresize}{\scriptsize}

\lstdefinelanguage{ioa}{
  basicstyle=\figuresize,
  keywordstyle=\bf \figuresize,
  identifierstyle=\it \figuresize,
  emphstyle=\tt \figuresize,
  mathescape=true,
  tabsize=20,
  sensitive=false,
  columns=fullflexible,
  keepspaces=false,
  flexiblecolumns=true,
  basewidth=0.05em,
  escapeinside={(*@}{@*)},
  moredelim=[il][\rm]{//},
  moredelim=[is][\sf \figuresize]{!}{!},
  moredelim=[is][\bf \figuresize]{*}{*},
  keywords={automaton,and, 
  	 choose,const,continue, components,
  	 discrete, do,
  	 eff, Eff, external,else, elseif, evolve, end,
  	 fi,for, forward, from,
  	 hidden,
  	 in,input,internal,if,invariant, initially, imports,
     let,
     or, output, operators, od, of,
     pre, Pre,
     return,
     such,satisfies, stop, signature, simulation, 
     trajectories,trajdef, transitions, that,then, type, types, to, tasks,
     variables, vocabulary, 
     when,where, with,while},
  emph={set, seq, tuple, map, array, enumeration},   
   literate=
        {(}{{$($}}1
        {)}{{$)$}}1
        {\\in}{{$\in\ $}}1
        {\\preceq}{{$\preceq\ $}}1
        {\\subset}{{$\subset\ $}}1
        {\\subseteq}{{$\subseteq\ $}}1
        {\\supset}{{$\supset\ $}}1
        {\\supseteq}{{$\supseteq\ $}}1
        {\\forall}{{$\forall$}}1
        {\\le}{{$\le\ $}}1
        {\\ge}{{$\ge\ $}}1
        {\\gets}{{$\gets\ $}}1
        {\\cup}{{$\cup\ $}}1
        {\\cap}{{$\cap\ $}}1
        {\\langle}{{$\langle$}}1
        {\\rangle}{{$\rangle$}}1
        {\\exists}{{$\exists\ $}}1
        {\\bot}{{$\bot$}}1
        {\\rip}{{$\rip$}}1
        {\\emptyset}{{$\emptyset$}}1
        {\\notin}{{$\notin\ $}}1
        {\\not\\exists}{{$\not\exists\ $}}1
        {\\ne}{{$\ne\ $}}1
        {\\to}{{$\to\ $}}1
        {\\implies}{{$\implies\ $}}1
        {<}{{$<\ $}}1
        {>}{{$>\ $}}1
        {=}{{$=\ $}}1
        {~}{{$\neg\ $}}1
        {|}{{$\mid$}}1
        {'}{{$^\prime$}}1
        {\\A}{{$\forall\ $}}1
        {\\E}{{$\exists\ $}}1
        {\\nE}{{$\nexists\ $}}1
        {\\/}{{$\vee\,$}}1
        {\\vee}{{$\vee\,$}}1
        {/\\}{{$\wedge\,$}}1
        {\\wedge}{{$\wedge\,$}}1
        {=>}{{$\Rightarrow\ $}}1
        {->}{{$\rightarrow\ $}}1
        {<=}{{$\Leftarrow\ $}}1
        {<-}{{$\leftarrow\ $}}1
        {~=}{{$\neq\ $}}1
        {\\U}{{$\cup\ $}}1
        {\\I}{{$\cap\ $}}1
        {|-}{{$\vdash\ $}}1
        {-|}{{$\dashv\ $}}1
        {<<}{{$\ll\ $}}2
        {>>}{{$\gg\ $}}2
        {||}{{$\|$}}1
        {[}{{$[$}}1
        {]}{{$\,]$}}1
        {[[}{{$\langle$}}1
        {]]]}{{$]\rangle$}}1
        {]]}{{$\rangle$}}1
        {<=>}{{$\Leftrightarrow\ $}}2
        {<->}{{$\leftrightarrow\ $}}2
        {(+)}{{$\oplus\ $}}1
        {(-)}{{$\ominus\ $}}1
        {_i}{{$_{i}$}}1
        {_j}{{$_{j}$}}1
        {_{i,j}}{{$_{i,j}$}}3
        {_{j,i}}{{$_{j,i}$}}3
        {_0}{{$_0$}}1
        {_1}{{$_1$}}1
        {_2}{{$_2$}}1
        {_n}{{$_n$}}1
        {_p}{{$_p$}}1
        {_k}{{$_n$}}1
        {-}{{$\ms{-}$}}1
        {@}{{}}0
        {\\delta}{{$\delta$}}1
        {\\R}{{$\R$}}1
        {\\Rplus}{{$\Rplus$}}1
        {\\N}{{$\N$}}1
        {\\times}{{$\times\ $}}1
        {\\tau}{{$\tau$}}1
        {\\alpha}{{$\alpha$}}1
        {\\beta}{{$\beta$}}1
        {\\gamma}{{$\gamma$}}1
        {\\ell}{{$\ell\ $}}1
        {--}{{$-\ $}}1
        {\\TT}{{\hspace{1.5em}}}3        
      }

\lstdefinelanguage{ioaNums}[]{ioa}
{
  numbers=left,
  numberstyle=\tiny,
  stepnumber=2,
  numbersep=4pt
}

\lstdefinelanguage{ioaNumsRight}[]{ioa}
{
  numbers=right,
  numberstyle=\tiny,
  stepnumber=2,
  numbersep=4pt
}

\lstnewenvironment{IOA}%
  {\lstset{language=IOA}}
  {}

\lstnewenvironment{IOANums}%
  {
  \if@firstcolumn
    \lstset{language=IOA, numbers=left, firstnumber=auto}
  \else
    \lstset{language=IOA, numbers=right, firstnumber=auto}
  \fi
  }
  {}

\lstnewenvironment{IOANumsRight}%
  {
    \lstset{language=IOA, numbers=right, firstnumber=auto}
  }
  {}

\newcommand{\linefigioa}[9]{

}

\lstdefinelanguage{ioaLang}{%
  basicstyle=\ttfamily\small,
  keywordstyle=\rmfamily\bfseries\small,
  identifierstyle=\small,
  keywords={assumes,automaton,axioms,backward,bounds,by,case,choose,components,const,d,det,discrete,do,eff,else,elseif,ensuring,enumeration,evolve,fi,fire,follow,for,forward,from,hidden,if,in,%
    input,initially,internal,invariant,let, local,od,of,output,pre,schedule,signature,so,%
    simulation,states,variables, tasks, stop,tasks,that,then,to,trajdef,trajectory,trajectories,transitions,tuple,type,union,urgent,uses,when,where,while,yield},
  literate=
        {\\in}{{$\in$}}1
        {\\preceq}{{$\preceq$}}1
        {\\subset}{{$\subset$}}1
        {\\subseteq}{{$\subseteq$}}1
        {\\supset}{{$\supset$}}1
        {\\supseteq}{{$\supseteq$}}1
        {\\rho}{{$\rho$}}1
        {\\infty}{{$\infty$}}1
        {<}{{$<$}}1
        {>}{{$>$}}1
        {=}{{$=$}}1
        {~}{{$\neg$}}1 
        {|}{{$\mid$}}1
        {'}{{$^\prime$}}1
        {\\A}{{$\forall$}}1 {\\E}{{$\exists$}}1
        {\\/}{{$\vee$}}1 {/\\}{{$\wedge$}}1 
        {=>}{{$\Rightarrow$}}1 
        {->}{{$\rightarrow$}}1 
        {<=}{{$\leq$}}1 {>=}{{$\geq$}}1 {~=}{{$\neq$}}1
        {\\U}{{$\cup$}}1 {\\I}{{$\cap$}}1
        {|-}{{$\vdash$}}1 {-|}{{$\dashv$}}1
        {<<}{{$\ll$}}2 {>>}{{$\gg$}}2
        {||}{{$\|$}}1
        {<=>}{{$\Leftrightarrow$}}2 
        {<->}{{$\leftrightarrow$}}2
        {(+)}{{$\oplus$}}1
        {(-)}{{$\ominus$}}1
}

\lstdefinelanguage{bigIOALang}{%
  basicstyle=\ttfamily,
  keywordstyle=\rmfamily\bfseries,
  identifierstyle=,
  keywords={assumes,automaton,axioms,backward,by,case,choose,components,const,%
    d,det,discrete,do,eff,else,elseif,ensuring,enumeration,evolve,fi,for,forward,from,hidden,if,in%
    input,initially,internal,invariant,local,od,of,output,pre,schedule,signature,so,%
    tasks, simulation,states,stop,tasks,that,then,to,trajdef,trajectories,transitions,tuple,type,union,urgent,uses,when,where,yield},
  literate=
        {\\in}{{$\in$}}1
        {\\preceq}{{$\preceq$}}1
        {\\subset}{{$\subset$}}1
        {\\subseteq}{{$\subseteq$}}1
        {\\supset}{{$\supset$}}1
        {\\supseteq}{{$\supseteq$}}1
        {<}{{$<$}}1
        {>}{{$>$}}1
        {=}{{$=$}}1
        {~}{{$\neg$}}1 
        {|}{{$\mid$}}1
        {'}{{$^\prime$}}1
        {\\A}{{$\forall$}}1 {\\E}{{$\exists$}}1
        {\\/}{{$\vee$}}1 {/\\}{{$\wedge$}}1 
        {=>}{{$\Rightarrow$}}1 
        {->}{{$\rightarrow$}}1 
        {<=}{{$\leq$}}1 {>=}{{$\geq$}}1 {~=}{{$\neq$}}1
        {\\U}{{$\cup$}}1 {\\I}{{$\cap$}}1
        {|-}{{$\vdash$}}1 {-|}{{$\dashv$}}1
        {<<}{{$\ll$}}2 {>>}{{$\gg$}}2
        {||}{{$\|$}}1
        {<=>}{{$\Leftrightarrow$}}2 
        {<->}{{$\leftrightarrow$}}2
        {(+)}{{$\oplus$}}1
        {(-)}{{$\ominus$}}1
}

\lstnewenvironment{BigIOA}%
  {\lstset{language=bigIOALang,basicstyle=\ttfamily}
   \csname lst@SetFirstLabel\endcsname}
  {\csname lst@SaveFirstLabel\endcsname\vspace{-4pt}\noindent}

\lstnewenvironment{SmallIOA}%
  {\lstset{language=ioaLang,basicstyle=\ttfamily\scriptsize}
   \csname lst@SetFirstLabel\endcsname}
  {\csname lst@SaveFirstLabel\endcsname\noindent}

\newlength{\bracklen}

\renewcommand{\arraystretch}{\defaultArraystretch}

\newcommand{\tri}[3]{\ensuremath{\mathit{#1}^\mathit{#2}_\mathit{#3}}}

\newcommand{\sugLocalVars}[2]{\ifthenelse{\equal{}{#2}}%
                             {\tri{localVars}{#1}{desug}}%
                             {\tri{localVars}{#1}{#2,desug}}}
\newcommand{\sugVars}[2]{\ifthenelse{\equal{}{#2}}%
                        {\tri{vars}{#1}{desug}}%
                        {\tri{vars}{#1}{#2,desug}}}

\newenvironment{subSyntax}{\begin{array}{l}}{\end{array}}

\newcommand{\ms}[1]{\ifmmode%
\mathord{\mathcode`-="702D\it #1\mathcode`\-="2200}\else%
$\mathord{\mathcode`-="702D\it #1\mathcode`\-="2200}$\fi}

\def\A{{\cal A}} % TA
\def\T{{\cal T}} % set of trajectories

\lstdefinelanguage{pvs}{
  basicstyle=\tt \figuresize,
  keywordstyle=\sc \figuresize,
  identifierstyle=\it \figuresize,
  emphstyle=\tt \figuresize,
  mathescape=true,
  tabsize=20,
  sensitive=false,
  columns=fullflexible,
  keepspaces=false,
  flexiblecolumns=true,
  basewidth=0.05em,
  moredelim=[il][\rm]{//},
  moredelim=[is][\sf \figuresize]{!}{!},
  moredelim=[is][\bf \figuresize]{*}{*},
  keywords={and, 
  	 begin,
  	 cases, const,
  	 do,
  	 external, else, exists, end, endcases, endif,
  	 fi,for, forall, from,
  	 hidden,
  	 in, if, importing,
     let, lambda, lemma,
     measure, 
     not,
     or, of,
     return, recursive,
     stop, 
     theory, that,then, type, types, type+, to, theorem,
     var,
     with,while},
  emph={nat, setof, sequence, eq, tuple, map, array, enumeration, bool, real, exp, nnreal, posreal},   
   literate=
        {(}{{$($}}1
        {)}{{$)$}}1
        {\\in}{{$\in\ $}}1
        {\\mapsto}{{$\rightarrow\ $}}1
        {\\preceq}{{$\preceq\ $}}1
        {\\subset}{{$\subset\ $}}1
        {\\subseteq}{{$\subseteq\ $}}1
        {\\supset}{{$\supset\ $}}1
        {\\supseteq}{{$\supseteq\ $}}1
        {\\forall}{{$\forall$}}1
        {\\le}{{$\le\ $}}1
        {\\ge}{{$\ge\ $}}1
        {\\gets}{{$\gets\ $}}1
        {\\cup}{{$\cup\ $}}1
        {\\cap}{{$\cap\ $}}1
        {\\langle}{{$\langle$}}1
        {\\rangle}{{$\rangle$}}1
        {\\exists}{{$\exists\ $}}1
        {\\bot}{{$\bot$}}1
        {\\rip}{{$\rip$}}1
        {\\emptyset}{{$\emptyset$}}1
        {\\notin}{{$\notin\ $}}1
        {\\not\\exists}{{$\not\exists\ $}}1
        {\\ne}{{$\ne\ $}}1
        {\\to}{{$\to\ $}}1
        {\\implies}{{$\implies\ $}}1
        {<}{{$<\ $}}1
        {>}{{$>\ $}}1
        {=}{{$=\ $}}1
        {~}{{$\neg\ $}}1
        {|}{{$\mid$}}1
        {'}{{$^\prime$}}1
        {\\A}{{$\forall\ $}}1
        {\\E}{{$\exists\ $}}1
        {\\/}{{$\vee\,$}}1
        {\\vee}{{$\vee\,$}}1
        {/\\}{{$\wedge\,$}}1
        {\\wedge}{{$\wedge\,$}}1
        {->}{{$\rightarrow\ $}}1
        {=>}{{$\Rightarrow\ $}}1
        {->}{{$\rightarrow\ $}}1
        {<=}{{$\Leftarrow\ $}}1
        {<-}{{$\leftarrow\ $}}1
        {~=}{{$\neq\ $}}1
        {\\U}{{$\cup\ $}}1
        {\\I}{{$\cap\ $}}1
        {|-}{{$\vdash\ $}}1
        {-|}{{$\dashv\ $}}1
        {<<}{{$\ll\ $}}2
        {>>}{{$\gg\ $}}2
        {||}{{$\|$}}1
        {[}{{$[$}}1
        {]}{{$\,]$}}1
        {[[}{{$\langle$}}1
        {]]]}{{$]\rangle$}}1
        {]]}{{$\rangle$}}1
        {<=>}{{$\Leftrightarrow\ $}}2
        {<->}{{$\leftrightarrow\ $}}2
        {(+)}{{$\oplus\ $}}1
        {(-)}{{$\ominus\ $}}1
        {_i}{{$_{i}$}}1
        {_j}{{$_{j}$}}1
        {_{i,j}}{{$_{i,j}$}}3
        {_{j,i}}{{$_{j,i}$}}3
        {_0}{{$_0$}}1
        {_1}{{$_1$}}1
        {_2}{{$_2$}}1
        {_n}{{$_n$}}1
        {_p}{{$_p$}}1
        {_k}{{$_n$}}1
        {-}{{$\ms{-}$}}1
        {@}{{}}0
        {\\delta}{{$\delta$}}1
        {\\R}{{$\R$}}1
        {\\Rplus}{{$\Rplus$}}1
        {\\N}{{$\N$}}1
        {\\times}{{$\times\ $}}1
        {\\tau}{{$\tau$}}1
        {\\alpha}{{$\alpha$}}1
        {\\beta}{{$\beta$}}1
        {\\gamma}{{$\gamma$}}1
        {\\ell}{{$\ell\ $}}1
        {--}{{$-\ $}}1
        {\\TT}{{\hspace{1.5em}}}3        
      }

\lstdefinelanguage{BigPVS}{
  basicstyle=\tt,
  keywordstyle=\sc,
  identifierstyle=\it,
  emphstyle=\tt ,
  mathescape=true,
  tabsize=20,
  sensitive=false,
  columns=fullflexible,
  keepspaces=false,
  flexiblecolumns=true,
  basewidth=0.05em,
  moredelim=[il][\rm]{//},
  moredelim=[is][\sf \figuresize]{!}{!},
  moredelim=[is][\bf \figuresize]{*}{*},
  keywords={and, 
  	 begin,
  	 cases, const,
  	 do, datatype,
  	 external, else, exists, end, endif, endcases,
  	 fi,for, forall, from,
  	 hidden,
  	 in, if, importing,
     let, lambda, lemma,
     measure,
     not,
     or, of,
     return, recursive,
     stop, 
     theory, that,then, type, types, type+, to, theorem,
     var,
     with,while},
  emph={nat, setof, sequence, eq, tuple, map, array, first, rest, add, enumeration, bool, real, posreal, nnreal},   
   literate=
        {(}{{$($}}1
        {)}{{$)$}}1
        {\\in}{{$\in\ $}}1
        {\\mapsto}{{$\rightarrow\ $}}1
        {\\preceq}{{$\preceq\ $}}1
        {\\subset}{{$\subset\ $}}1
        {\\subseteq}{{$\subseteq\ $}}1
        {\\supset}{{$\supset\ $}}1
        {\\supseteq}{{$\supseteq\ $}}1
        {\\forall}{{$\forall$}}1
        {\\le}{{$\le\ $}}1
        {\\ge}{{$\ge\ $}}1
        {\\gets}{{$\gets\ $}}1
        {\\cup}{{$\cup\ $}}1
        {\\cap}{{$\cap\ $}}1
        {\\langle}{{$\langle$}}1
        {\\rangle}{{$\rangle$}}1
        {\\exists}{{$\exists\ $}}1
        {\\bot}{{$\bot$}}1
        {\\rip}{{$\rip$}}1
        {\\emptyset}{{$\emptyset$}}1
        {\\notin}{{$\notin\ $}}1
        {\\not\\exists}{{$\not\exists\ $}}1
        {\\ne}{{$\ne\ $}}1
        {\\to}{{$\to\ $}}1
        {\\implies}{{$\implies\ $}}1
        {<}{{$<\ $}}1
        {>}{{$>\ $}}1
        {=}{{$=\ $}}1
        {~}{{$\neg\ $}}1
        {|}{{$\mid$}}1
        {'}{{$^\prime$}}1
        {\\A}{{$\forall\ $}}1
        {\\E}{{$\exists\ $}}1
        {\\/}{{$\vee\,$}}1
        {\\vee}{{$\vee\,$}}1
        {/\\}{{$\wedge\,$}}1
        {\\wedge}{{$\wedge\,$}}1
        {->}{{$\rightarrow\ $}}1
        {=>}{{$\Rightarrow\ $}}1
        {->}{{$\rightarrow\ $}}1
        {<=}{{$\Leftarrow\ $}}1
        {<-}{{$\leftarrow\ $}}1
        {~=}{{$\neq\ $}}1
        {\\U}{{$\cup\ $}}1
        {\\I}{{$\cap\ $}}1
        {|-}{{$\vdash\ $}}1
        {-|}{{$\dashv\ $}}1
        {<<}{{$\ll\ $}}2
        {>>}{{$\gg\ $}}2
        {||}{{$\|$}}1
        {[}{{$[$}}1
        {]}{{$\,]$}}1
        {[[}{{$\langle$}}1
        {]]]}{{$]\rangle$}}1
        {]]}{{$\rangle$}}1
        {<=>}{{$\Leftrightarrow\ $}}2
        {<->}{{$\leftrightarrow\ $}}2
        {(+)}{{$\oplus\ $}}1
        {(-)}{{$\ominus\ $}}1
        {_i}{{$_{i}$}}1
        {_j}{{$_{j}$}}1
        {_{i,j}}{{$_{i,j}$}}3
        {_{j,i}}{{$_{j,i}$}}3
        {_0}{{$_0$}}1
        {_1}{{$_1$}}1
        {_2}{{$_2$}}1
        {_n}{{$_n$}}1
        {_p}{{$_p$}}1
        {_k}{{$_n$}}1
        {-}{{$\ms{-}$}}1
        {@}{{}}0
        {\\delta}{{$\delta$}}1
        {\\R}{{$\R$}}1
        {\\Rplus}{{$\Rplus$}}1
        {\\N}{{$\N$}}1
        {\\times}{{$\times\ $}}1
        {\\tau}{{$\tau$}}1
        {\\alpha}{{$\alpha$}}1
        {\\beta}{{$\beta$}}1
        {\\gamma}{{$\gamma$}}1
        {\\ell}{{$\ell\ $}}1
        {--}{{$-\ $}}1
        {\\TT}{{\hspace{1.5em}}}3        
      }

\lstdefinelanguage{pvsNums}[]{pvs}
{
  numbers=left,
  numberstyle=\tiny,
  stepnumber=2,
  numbersep=4pt
}

\lstdefinelanguage{pvsNumsRight}[]{pvs}
{
  numbers=right,
  numberstyle=\tiny,
  stepnumber=2,
  numbersep=4pt
}

\lstnewenvironment{BigPVS}%
  {\lstset{language=BigPVS}}
  {}

\lstnewenvironment{PVSNums}%
  {
  \if@firstcolumn
    \lstset{language=pvs, numbers=left, firstnumber=auto}
  \else
    \lstset{language=pvs, numbers=right, firstnumber=auto}
  \fi
  }
  {}

\lstnewenvironment{PVSNumsRight}%
  {
    \lstset{language=pvs, numbers=right, firstnumber=auto}
  }
  {}

\newcommand{\linefigpvs}[9]{

}

\lstdefinelanguage{pvsproof}{
  basicstyle=\tt \figuresize,
  mathescape=true,
  tabsize=4,
  sensitive=false,
  columns=fullflexible,
  keepspaces=false,
  flexiblecolumns=true,
  basewidth=0.05em,
}

\newcommand{\tuple}[1]{\left\langle#1\right\rangle}

\newcommand{\ceil}[1]{\left\lceil#1\right\rceil}

\def\Loc{\act{Loc}}

\newcommand{\toolQBMC}{QBMC\xspace}

\newcommand{\ds}[1]{\left\llbracket#1\right\rrbracket}	%denotation of states

\def\N{\act{N}}

\def\Loc{\act{Loc}}

\newcommand{\val}[1]{{\mathit{val}(#1)}}

\newcommand{\localvar}[2]{{{#1_{#2}}}}
\newcommand{\localvardot}[2]{{\dot{#1}_{#2}}}

\def\xi{\localvar{x}{i}}

\def\qi{\localvar{q}{i}}
\def\qk{\localvar{q}{k}}

\def\lock{\localvar{\ell}{k}}
\def\valk{\localvar{\val}{k}}
\def\loco{\localvar{\ell}{0}}
\def\valo{\localvar{\val}{0}}

\def\qiplusone{\localvar{q}{i + 1}}

\def\Ti{\localvar{T}{i}}

\def\dotline{...}

\def\iniset{I}
\def\badset{P}
\def\BMC{\Phi(k)}
\def\BMCquantifier{\Omega(k)}
\def\BMCquantifierthree{\Omega(3)}
\def\Fischersafe{FS\xspace}
\def\Fischerunsafe{FU\xspace}
\def\Fischersafetwo{FS-2\xspace} % Taylor2Luan: probably this would be simpler with a \newcommand that takes an argument, e.g. \fischerunsafe{2}
\def\Fischerunsafetwo{FU-2\xspace}
\def\Fischersafethree{FS-3\xspace}
\def\Fischerunsafethree{FU-3\xspace}
\def\Fischersafefour{FS-4\xspace}
\def\Fischerunsafefour{FU-4\xspace}
\def\Fischersafefive{FS-5\xspace}
\def\Fischerunsafefive{FU-5\xspace}

\def\reach{{\sf Reach}}

\def\locrem{\act{rem}}
\def\loctry{\act{try}}
\def\locwait{\act{wait}}
\def\loccs{\act{cs}}

\newcommand{\Invset}{{\mathit{Inv}}}
\newcommand{\Initset}{{\mathit{Init}}}
\newcommand{\Flowset}{{\mathit{Flow}}}
\newcommand{\flowrate}{{\mathit{f}}}
\newcommand{\Locset}{{\mathit{Loc}}}
\newcommand{\Varset}{{\mathit{Var}}}
\newcommand{\Transset}{{\mathit{Trans}}}
\newcommand{\Transrelation}{T}
\newcommand{\DisTrans}{D}
\newcommand{\Trajectory}{\T}
\newcommand{\guard}{g}
\newcommand{\reset}{u}

\def\Xi{\mathit{X_i}}

\def\q{q}
\def\val{\mathit{v}}

\def\AutomatonH{\H}
\newcommand{\Real}{\mathbb{R}}
\newcommand{\nnReal}{{\nnnum \Real}} 
\newcommand{\Realn}{\mathbb{R}^n}
\newcommand{\Flowfunction}{\Delta}
\newcommand{\talpha}{t_{\alpha}}
\newcommand{\tbeta}{t_{\beta}}
\newcommand{\locone}{\ell\mathit{oc_{1}}}
\newcommand{\loctwo}{\ell\mathit{oc_{2}}}

\newcommand{\eglabel}[1]{\label{def:#1}}
\newcommand{\egref}[1]{Example~\ref{def:#1}}

%% file: abstract.tex
%SAT Modulo Theories (SMT) solvers included 
%
%Boolean Satisfiability (SAT) solvers and Quantified Boolean Formulas (QBF)  
%
%
Satisfiability Modulo Theories (SMT) solvers have been successfully applied to solve many problems in formal verification such as bounded model checking (BMC) for many classes of systems from integrated circuits to cyber-physical systems. Typically, BMC is performed by checking satisfiability of a possibly long, but quantifier-free formula. However, BMC problems can naturally be encoded as quantified formulas over the number of BMC steps. In this approach, we then use decision procedures supporting quantifiers to check satisfiability of these quantified formulas. This approach has previously been applied to perform BMC using a Quantified Boolean Formula (QBF) encoding for purely discrete systems, and then discharges the QBF checks using QBF solvers. In this paper, we present a new quantified encoding of BMC for rectangular hybrid automata (RHA), which requires using more general logics due to the real (dense) time and real-valued state variables modeling continuous states. We have implemented a preliminary experimental prototype of the method using the HyST model transformation tool to generate the quantified BMC (QBMC) queries for the Z3 SMT solver. We describe experimental results on several timed and hybrid automata benchmarks, such as the Fischer and Lynch-Shavit mutual exclusion algorithms. We compare our approach to quantifier-free BMC approaches, such as those in the dReach tool that uses the dReal SMT solver, and the HyComp tool built on top of nuXmv that uses the MathSAT SMT solver. Based on our promising experimental results, QBMC may in the future be an effective and scalable analysis approach for RHA and other classes of hybrid automata as further improvements are made in quantifier handling in SMT solvers such as Z3.

%% file: intro.tex
\section{Introduction}
\seclabel{intro}
%\vspace{-0.25em}
%
Boolean Satisfiability (SAT) is the canonical NP-complete problem and is to determine if a given Boolean formula is satisfiable, \ie check if there exists an assignment of values to variables where the formula is true.
A Boolean formula is given in Conjunctive Normal Form (CNF), that is, a conjunction of clauses, each of which is a disjunction of literals.
Satisfiability Modulo Theories (SMT) is a generalization of SAT, where literals are interpreted with respect to a background theory (\eg linear real arithmetic, nonlinear integer arithmetic, bit-vectors, etc.).
%
%Recently, SMT-based techniques have been developed to formally verify hybrid systems~\cite{cimatti2013smt, cimatti2012quantifier, gao2013dreal}.
%
%,mangassarian2010tc,jussila2007compressing

SMT-based techniques have been developed to formally verify hybrid systems~\cite{eggers2008atva,gao2013fmcad,johnson2014formats,cimatti2015tacas,shmarov2017smt,bae2017modular}.
Typically, these SMT-based methods are used in bounded model checking (BMC), which is to check for a transition system $A$ and a specification $P$ whether $I(V_0) \wedge \bigwedge_{i=0}^{k-1} T(V_i,V_{i+1}) \wedge (\bigvee_{i=0}^{k} P(V_i))$ is satisfiable.
Here, $I(V_0)$ encodes an initial set of states over a set of variables $V_0$, $\Transrelation(V_i,V_{i+1})$ represents the transition relation from iteration $i$ to $i+1$ over sets of variables $V_i$ and $V_{i+1}$, and $P(V_i)$ encodes the specification at step $i$.
An SMT solver either returns SAT if there is a sequence of states leading the transition system $A$ from a state in $I$ to a state in $P$, or UNSAT if a state in $P$ can not be reached in $k$ steps. 
In principle, BMC is complete as it can prove that no bad state can be reached if a large enough bound k is used. However, as k increases, SMT-based BMC may require excessive memory due to the underlying complex combination theories in a SMT solver. To increase scalability, it is essential to reduce the memory usage while performing SMT-based BMC for large hybrid systems.

Hybrid automaton is a modeling formalism used to verify dynamical systems including both continuous states and dynamics as well as discrete states and transitions.
Examples of systems naturally modeled by hybrid automata arise in the interaction of physical plants and software controllers in real-time systems and cyber-physical systems (CPS).
In essence, hybrid automata augment finite state machines with a set of real-valued variables that evolve continuously over intervals of real time.
In hybrid automata, a transition relation $\Transrelation = \DisTrans \cup \Trajectory$ encodes both discrete transitions $\DisTrans$ and continuous trajectories $\Trajectory$ over intervals of real-time. 
Rectangular hybrid automata (RHA) are a special class of hybrid automata with continuous dynamics described by rectangular differential inclusions and where all other quantities (guard conditions, invariants, resets, etc.) of the automata are linear inequalities over constants~\cite{henzinger1996lics,johnson2012forte}.
Sets of states, as well as discrete transitions and continuous trajectories of RHA, can be symbolically represented by SMT formulas over real and Boolean variables. 

Depending on the underlying logics supported, SMT solvers may or may not support quantifiers.
While quantifiers may make the language more expressive, they often increase the complexity of computations like checking satisfiability and may also lead to undecidability.
Techniques allowing quantifiers, such as in quantified Boolean formula (QBF) solvers, have been developed for the BMC of purely discrete systems, such as finite state machines~\cite{jussila2007compressing,mangassarian2010tc}. 
However, to the best of our knowledge, there has been no effort to develop quantified BMC (QBMC) methods for timed or hybrid automata, which we develop in this paper.
Of course, this is partially because the underlying SMT solver requires support for complex combination theories and efficient algorithms to check quantified formulas, which until recently, were either not available or not scalable. 
%
%

%\vspace{1em}
%\noindent
%\paragraph*{Contributions.}
%
In this paper, we propose a SMT-based verification technique that encodes the BMC problem for timed automata and RHA in a quantified form, which we call QBMC (quantified bounded model checking). 
As the logic encoding requires some finite sort for the discrete states (such as a enumerated type or bit-vectors) and reals for the continuous states and trajectories, we uses LRABV (linear real arithmetic with bit-vectors) for encoding QBMC for timed automata and RHA. We note that general hybrid automata would need NRABV (nonlinear real arithmetic with bit-vectors) or beyond, such as those whose solutions involve special (transcendental) functions like sin, cos, exp, etc.
While none of these logics are officially supported in the SMT-LIB2 standard as of the time of this writing~\cite{BarFTRR17}, several solvers do have unofficial support for this combination theory, such as the latest versions of Z3, which is the SMT solver used in this paper~\cite{demoura2008tacas}.

For implementation, we take hybrid automata in the SpaceEx format~\cite{frehse2011cav}, which are then translated to the QBMC encoding sequence proposed in this paper using the HyST model transformation tool~\cite{bak2015hscc}. HyST allows the same model to be analyzed simultaneously in several hybrid systems analysis tools, so it is convenient to compare the performance of our proposed approach with existing works. 
The QBMC is performed by querying the Z3 SMT solver via its Python API and using its quantifier-handling procedures~\cite{demoura2008tacas}.
We present preliminary experimental results where the QBMC approach and Z3 perform competitively to (a) the dReach tool that performs BMC using an SMT check by querying the dReal $\delta$-decidable SMT solver~\cite{gao2012lics,gao2013fmcad,kong2015dreach}, and (b) the HyComp tool built on top of nuXmv that uses the MathSAT SMT solver~\cite{cimatti2015hycomp}.
The examples include standard ones such as Fischer and Lynch-Shavit mutual exclusion, as well as an illustrative example to describe the encoding.
Overall, the main contribution of this paper is the first encoding of BMC as a quantified problem for RHA that has less memory requirement and competitive execution time when compared to state-of-the-art SMT solvers.
Our results subsume the case for timed automata, as RHA are more expressive than timed automata, and we note this is also the first QBMC approach for timed automata.
%
%Another contribution is the implementation and experimental evaluation of the method using HyST and Z3.

%% file: rw.tex
\section{Related Work}
\seclabel{rw}
When defining the semantics of hybrid automata, first-order or higher logic is typically used and quantifiers typically show up in several places.
Existential quantifiers over reals are used to specify that some amount of real time may elapse in a given location of the hybrid automaton.
Universal quantifiers over reals representing real time are used to construct invariants that are enforced at all times, while in a given location of the hybrid automaton; otherwise real time is not allowed to advance, and a discrete transition must be taken, if any are enabled based on the current state and guards of the transitions.
Alternative approaches to the one described in this paper have previously been developed, where the universal quantifiers used to define invariant semantics are explicitly removed from the SMT expressions to create quantifier-free formulas.
That allows the use of existing SMT-based procedures and avoids quantifier-elimination and other quantifier-handling procedures~\cite{johnson2012forte,cimatti2013fmsd}.
We note those approaches do not use quantifiers on the number of steps $k \geq 0$ in the BMC computation, which we do in this paper.
Specifically, we suggest that effectively encoding the BMC problem in a quantified form over the number of steps $k$ may provide a more scalable approach in the future as quantifier handling procedures are improved in the underlying solvers.
We accomplish this by extending existing results for the BMC of discrete systems with QBF solvers~\cite{mangassarian2010tc} to timed and hybrid automata, specifically RHA.

Typical approaches that analyze timed and hybrid automata use symbolic representations of states such as difference bound matrices (DBMs) to represent clock regions in Uppaal~\cite{bengtsson1995}, HybridSAL~\cite{tiwari2012cav}, or polyhedra in HyTech~\cite{henzinger1997sttt}. 
Several other formal verification tools for hybrid automata focus on performing reachability computations, and overapproximate the set of reachable states using various data structures to symbolically represent geometric sets of states, such as Taylor models in Flow*~\cite{chen2013cav} and support functions in SpaceEx~\cite{frehse2011cav}.
Reachability analysis tools like Flow* and SpaceEx focus on computing reachable states, although there is a direct equivalence between time-bounded reachability computations and BMC.
% and PHAVer~\cite{frehse2008sttt}.

Several SMT-based approaches can verify properties of timed and hybrid automata.
%
%For timed automata, BMC and IC3 algorithms have been developed that query SMT solvers~\cite{kindermann2012forte}.
%kindermann2012formats
%
dReal is an SMT-solver for first-order logic formulas over the reals, and uses a $\delta$-complete decision procedure~\cite{gao2012lics}.
dReach is a BMC tool that queries dReal to check satisfiability of SMT formulas encoding the transitions and trajectories for hybrid automata~\cite{gao2013fmcad}.
HyComp is a verification tool for networks (parallel compositions) of hybrid automata with polynomial and other dynamics~\cite{cimatti2015tacas} and is built on top of nuXmv~\cite{cavada2014cav}.
HyComp supports several verification modes, including a BMC analysis mode, $k$-induction, and IC3.
For $k$-induction and IC3, HyComp may perform unbounded model checking, but in the BMC mode, it also allows a limit on the number of steps, and also encodes the semantics of the network of hybrid automata's transition relation and trajectories. In this paper, we will compare our QBMC approach to dReach and HyComp in terms of  performing BMC for hybrid systems.

A very closely related approach to this paper also encodes BMC problems for timed automata using quantified formulas, but this quantification is to encode unknown or incomplete components, and is not a quantification over the BMC length~\cite{miller2011mtv}. %ATMOC is an SMT-based model checker for timed automata.
Passel is a parameterized verification tool for networks of RHA that may prove properties regardless of the number $N$ of automata in the network~\cite{johnson2012forte}.
Passel implements an extension to hybrid automata of the invisible invariants approach for parameterized verification, and consists of an invariant synthesis procedure~\cite{johnson2013infotech} that relies on reachability computations~\cite{johnson2014formats}.
Passel encodes the semantics of networks of hybrid automata as SMT formulas and checks satisfiability and validity using the Z3 SMT solver.
When performing reachability computations, Passel makes use of quantifier elimination procedures over the reals and bit-vectors~\cite{johnson2014formats}. 
MCMT is another SMT-based verification tool equipped with an extensive quantifier instantiation to verify the safety properties of parametrized systems~\cite{ghilardi2010mcmt}. 
In contrast, our proposed approach straightforwardly deals with quantified SMT formulas.

%\commenttaylor{Others: HybridSAL~\cite{tiwari2012cav}, and Kindermann's work~\cite{kindermann2012forte,kindermann2012formats} and compare to Uppaal that uses difference bound matrices (DBMs) to symbolically represent clock regions~\cite{bengtsson1995}.}

%% file: preliminary.tex
\section{Preliminaries}

\def\VarsetShort{V}
In this section, we introduce the preliminaries that are needed for this work. We first define a hybrid automaton model, discuss its semantics and safety specification. Then we present the traditional encoding (quantifier-free) of the BMC for hybrid automata.

\subsection{Hybrid Automata}
%\vspace{-0.25em}
%\paragraph*{Preliminaries}
\noindent
{\bf Syntax.} A hybrid automaton is essentially a finite state machine extended with a set of real-valued variables that evolve continuously over intervals of real-time.
%
%\paragraph*{Syntax.}
%
The syntactic structure of a hybrid automaton is formally defined as follows.
%
%\begin{definition}
%\deflabel{Automaton}
%
A hybrid automaton $\AutomatonH$ is a tuple, $\AutomatonH$ $\deq$ $\langle$$\Locset$, $\Varset$, $\Invset$, $\Flowset$, $\Transset$, $\Initset$$\rangle$, with the components as follows.
\begin{itemize}
%\begin{inparaenum}[(a)]
%
\item $\Locset$ is a finite set of discrete locations.
%
%\item $\mathtt{Lab}$: a finite set of synchronization labels,
\item $\Varset$ is a finite set of $n$ continuous, real-valued variables, and $\Q \deq \Locset \times \Realn$ is the state-space.
\item $\Invset$ is a finite set of invariants, one for each discrete location, and for each location $\ell \in \Locset$, $\Invset(\ell) \subseteq \Realn$.
\item $\Flowset$ is a finite set of ordinary differential inclusions, one for each continuous variable $x \in \Varset$, and $\Flowset(\ell, x) \subseteq \Realn$ describes the continuous dynamics in each location $\ell \in \Locset$.
\item $\Transset$ is a finite set of transitions between locations.
Each transition is a tuple $\tau \deq \tuple{\ell, \ell', \guard, \reset}$, where $\ell$ is a source location and $\ell'$ is a target location that may be taken when a guard condition $\guard$ is satisfied, and the post-state is updated by an update map $\reset$.
%
%Thus, $\forall x \in \mathtt{Grd}$, $x' \in \mathtt{Rst}$, $\exists q, q' \in \Q$, if $\mathtt{Grd} \subseteq \R^n$ and $\mathtt{Rst}(x) \subseteq \R^n$, then $q(l, x) \to q'(l', x')$.
%
\item $\Initset$ is an initial condition, which consists of a set of locations in $\Locset$ and a formula over $\Varset$, so that $\Initset \subseteq \Q$.
\end{itemize}
%\end{inparaenum}
%\vspace{1em}%
%\end{definition}
%
%For LHA, all the expressions appearing in invariants, guards, and updates must be boolean combinations of linear expressions, and the flows are rectangular differential inclusions ($\dot{x} \in [a, b]$ for $a \leq b$)~\cite{henzinger1996lics}.
%

For RHA, all the expressions appearing in invariants, guards, and updates must be boolean combinations of constant inequalities, and the flows are rectangular differential inclusions e.g., $\dot{x} \in [a, b]$ for $a \leq b$. %~\cite{henzinger1996lics}
Timed automata is the important subclass of RHA in which every continuous variable is a precise clock e.g., $\dot{x} = 1$.  
We use the dot (.) notation to refer to different components of tuples \eg $\AutomatonH.\Invset$ refers to the invariants of automaton $\AutomatonH$ and $\tau.\guard$ refers to the guard of a transition $\tau$.
If clear from context, we drop $\AutomatonH$ and $\tau$ and refer to the individual components of the tuple.

%
%\commenttaylor{Probably want to say we have a special variable $q$ that takes values in $\Locset$ so we can write out the BMC conveniently (possible without this, but maybe less clear). Also want to restrict the differential inclusions, invariants, resets, etc. to be from some class following the definition. E.g., timed would be $\dot{x} = 1$ and all resets, guards, etc. are constants, etc. The method we describe could in theory be applied up to ODEs with polynomial solutions, as these can be described exactly in SMT solvers. Beyond those, we'd have to start working with abstractions for decidability reasons (e.g., if the solutions involved exponentials or other transcendentals).}

\vspace{1em}
\noindent
{\bf Semantics.}
%\paragraph*{Semantics.}
%
The semantics of a hybrid automaton $\AutomatonH$ are defined in terms of executions, which are sequences of states.
A \emph{state} $\q$ of $\AutomatonH$ is a tuple $\q \deq \tuple{\ell, \val}$, where $\ell \in \Locset$ is a location, and $\val \in \Realn$ is a valuation of all variables in $\Varset$.
Formally, for a set of variables $\Varset$, a \emph{valuation} is a function mapping each $x \in \Varset$ to a point in its type---here, $\Real$.
%
%$\forall x \in \Varset$, $\val.x \in \Real$
%
The state-space $\Q$ is the set of all states of $\AutomatonH$.
Updates of states are described by a transition relation $\Transrelation \subseteq \Q \times \Q$. 
For a transition $\tuple{\q, \q'} \in \Transrelation$ where $\q \deq \tuple{\ell, \val}$ and $\q' \deq \tuple{\ell', \val'}$, we denote  $\q \rightarrow \q' \in \Transrelation$ as the transition between the current state $\q$ and the next state $\q'$.
The transition relation $\Transrelation$ is partitioned into disjoint sets of discrete transitions and continuous trajectories that respectively describe the discrete and continuous behaviors of the automaton.
Thus, $\Transrelation \deq \DisTrans \cup \Trajectory$, where:
\begin{inparaenum}[(a)]
	\item $\DisTrans \subseteq \Q \times \Q$ is the set of discrete transitions that describe instantaneous updates of state,
	\item $\Trajectory \subseteq \Q \times \Q$ is the set of continuous trajectories that describe updates of state over real time intervals.
\end{inparaenum}

\vspace{1em}
\emph{Discrete transitions}.
A discrete transition $\q \rightarrow \q' \in \DisTrans$ models an instantaneous update from the current state $\q$ to the next state $\q'$.
There is a discrete transition $\q \rightarrow \q' \in \DisTrans$ if and only if (iff):
$\exists \tau \in \Transset : \q.\val \models \act{\tau}.\guard \wedge \q'.\val' \models \act{\tau}.\reset$, where $\act{\tau}.\guard$, and $\act{\tau}.\reset$ are the guard condition and the update map of the discrete transition $\tau$, respectively.
%
%%$Therefore, discrete transitions can be used to switch the current state to the next state iff the guard conditions and the next state invariant are satisfied.
%
%$\guard(\act{\tau})$, and $\reset(\act{\tau})$

\vspace{1em}
\emph{Continuous trajectories}.
A continuous trajectory $\q \rightarrow \q' \in \Trajectory$  models the update of state $\q$ to $\q'$ over an interval of real time.
The set-valued function $\Flowfunction$ returns a set of states and is defined as: $\Flowfunction(\q.\ell, \q.v, x, t) \in \q.v.x + \int_{\delta = t_0}^{t} \flowrate(\q.\ell, x) d\delta$, where $\flowrate \in \Flowset$ is a flow rate---a formula over $\Varset \cup \dot{\Varset}$ that describes the evolution of a real variables $x \in \Varset$ over a real time interval $J = [t_0, t]$--- and $\q.v.x$ is the value of continuous variable $x$ of the state $\q$ at $t = t_0$.
Then, there is a trajectory $\q \rightarrow \q' \in \Trajectory$ iff: 
$\exists \talpha \in \nnReal \ \forall \tbeta \in \nnReal \ \exists \ell \in \Locset \ : $ $\tbeta \leq \talpha \ \wedge \ \Flowfunction(\q.\ell, \q.v, \Varset, \tbeta) \models \Invset(\ell) \ \wedge \ \q'.\val'.\Varset \in \Flowfunction(\q.\ell, \q.v, \Varset, \talpha)$.
%
%$\exists \talpha \in \nnReal \ \forall \tbeta \in \nnReal \ \exists \ell \in \LocType \ : $ $\tbeta \leq \talpha \ \wedge \ \Flowfunction(\q.\ell, \q.v, \Varset, \tbeta) \models \Invset(\ell) \ \wedge \ \q'.\VarSet \in \Flowfunction(\q.\ell, \q.v, \Varset, \alpha)$.
%
For each real variable $x$, $\q.\val.x$ must evolve to the valuation $\q'.\val'.x$ at precisely time $\talpha$ and corresponding to the flow rate of $x$ in location $\ell$.
Additionally, all states along the trajectory must satisfy the invariant $\Invset(\ell)$ \ie at every point in the interval of real time $\tbeta \leq \talpha$.

\vspace{1em}
\emph{Executions}.
An \emph{execution} of $\AutomatonH$ is a sequence $\pi \deq \q_0 \rightarrow \q_1 \rightarrow \q_2 \rightarrow \dotline$, such that:
\begin{inparaenum}[(a)]
	\item $\q_0 \in \Initset$ is an initial state, and
	\item either $\qi \rightarrow \qiplusone \in \DisTrans$ is a discrete transition or $\qi \rightarrow \qiplusone \in \Trajectory$ is a continuous trajectory for each consecutive pair of states in the sequence $\pi$.
\end{inparaenum}
%
%
%When necessary to specify the type of transition relations, we denote $\qi \xrightarrow{\Traji} \qiplusone$ and $\qi \xrightarrow{\DisTrani} \qiplusone$ as the discrete transition and the continuous trajectory of a consecutive pair of states $\tuple{\qi, \qiplusone}$, respectively.
%
A state $\qk \deq \tuple{\lock,\valk}$ is \emph{reachable} from initial state $\q_0 \deq \tuple{\loco,\valo} \in \Initset$ iff there exists a finite execution $\pi \deq \q_0 \rightarrow \q_1 \rightarrow \dotline \rightarrow \qk$.

\vspace{1em}
\noindent
{\bf Safety specifications.}
In this paper, we develop the QBMC procedure to check whether safety properties of hybrid automata are satisfied up to iteration $k$.
A \emph{safety specification} $\phi$ is a formula over $\Locset$ and $\Varset$ that describes a set of states $\ds{\phi} \subseteq \Q$, where $\ds{\cdot}$ is the set of states satisfying $\phi$.
For an automaton $\AutomatonH$ and a safety specification $\phi$, the automaton satisfies the specification, denoted $\AutomatonH \models \phi$, iff for every execution $\pi$, for every state $\q_0, \q_1, \ldots, \q_k$ in the execution $\pi$, we have $\pi.q_k \in \ds{\phi}$.
If $\AutomatonH \models \phi$ for every $i \in \{0, \ldots, k\}$, then the system is safe up to iteration $k$.
If $\AutomatonH \models \phi$ for any $k$, then the system is safe.
For a safety specification $\phi$, a \emph{counterexample} is an execution $\pi$ where some state $\q \in \pi$ violates $\phi$, \ie $\q \not\models \phi$, or equivalently, $\q \notin \ds{\phi}$.

\subsection{Quantifier-Free BMC for Hybrid Automata}

BMC has been used widely in verification and falsification of safety and liveness properties of various classes of systems, from finite state machines to hybrid automata.
The key idea is to search for a counterexample execution whose length is bounded by a number of steps $k$.
In other words, BMC will explore all executions from any initial state of a system to detect whether there is a way to reach a bad state that violates a given property (or to find a loop in the case of liveness).
Then this path is considered as a counterexample to the property that may help the user to debug the system. 
For finite state systems, BMC can be encoded as a propositional formula to be checked as satisfiable or unsatisfiable using a Boolean SAT solver.
For hybrid automata, BMC can be encoded as a formula over reals and finite sorts (such as Booleans, bit-vectors, or enumerated types).
In this paper, we focus only on hybrid automata with rectangular differential inclusion dynamics ($\dot{x} \in [a, b]$ for real constants $a \leq b$), and for this class of automata, the formulas are within linear real arithmetic (LRA).
Before presenting the QBMC approach in \secref{algorithm}, which is the main contribution of this paper, we first illustrate BMC for hybrid automata using the traditional quantifier-free encoding.

Let $\badset$ be a set of given specifications of the hybrid automata, the BMC problem will determine whether a specification $\badset(\qk) \in \badset$ is safe after $k$ steps, and it is:
%\wedge (\bigvee_{i=0}^{k-1} P(s_i))
\begin{equation} 
 %\BMC = \exists \q_0, \q_1,\dotline,\qkminusone \mid \iniset(\q_0) \wedge \bigwedge_{i = 0}^{k-1}{\Tiplusone} \wedge \badset(\qk).
%
%\BMC = \exists \q_0, \q_1,\dotline,\qk \mid \iniset(\q_0) \wedge \bigwedge_{i = 0}^{k-1}{\T(q_i, q_{i+1}} \wedge \bigvee_{i = 0}^{k}{\badset(\qi)}.
%
%\BMC = \exists \q_0, \q_1,\dotline,\qk \mid \iniset_0 \wedge \bigwedge_{i = 0}^{k-1}{\Tiplusone} \wedge (\bigvee_{i = 0}^{k}{\badset_i}).
%\BMC = \iniset_0 \wedge \bigwedge_{i = 0}^{k-1}{\Tiplusone} \wedge (\bigvee_{i = 0}^{k}{\badset_i}),
%\BMC \deq \iniset(\VarsetShort_0) \wedge \bigwedge_{i = 0}^{k-1}{\Ti(\VarsetShort, \VarsetShort')} \wedge (\bigvee_{i = 0}^{k}{\badset(\VarsetShort_i)}),
\BMC \deq \iniset(\VarsetShort_0) \wedge \bigwedge_{i = 0}^{k-1}{\Ti(\VarsetShort_i, \VarsetShort_{i+1})} \wedge (\bigvee_{i = 0}^{k}{\badset(\VarsetShort_i)}),
\eqlabel{bmc_smt}
\end{equation}
where $\VarsetShort_i$ corresponds to the set of variables $\Varset$ of the automaton $\AutomatonH$ appropriately renamed.
%
%For example, $\VarsetShort_i$ contains of every variable $v \in \Varset$ syntactically renamed to $v_i$, etc., and $\VarsetShort'$ consists of primed variables, e.g., $v'$ for each $v \in \Varset$.
%
In~\eqref{bmc_smt}, $\iniset(\VarsetShort_0)$ encodes the initial set of states, $\Ti(\VarsetShort_i, \VarsetShort_{i+1})$ encodes the transition between consecutive pairs of sets of states, and $\badset(\VarsetShort_i)$ is a safety specification at iteration $i$.
We note that the sets of variables $\VarsetShort_i$ for each iteration $i$ are implicitly existentially quantified, \eg we could equivalently prefix $\exists \VarsetShort_0, \VarsetShort_1, \ldots, \VarsetShort_k$.
We drop the sets of variables for a shorter notation, \eg~\eqref{bmc_smt} is equivalent to $\iniset_0 \wedge \bigwedge_{i = 0}^{k-1}{\Ti} \wedge (\bigvee_{i = 0}^{k}{\badset_i})$.

\begin{figure}[t!]%
	%\vspace{-0.5em}%
	\centering%
	\begin{adjustbox}{max size={0.75\columnwidth}{0.75\textheight}}%
	\begin{tikzpicture}[-,>=stealth',shorten >=1pt,auto,node distance=8cm,main node/.style={circle,fill=black!20,draw}]
		\tikzset{every state/.style={minimum size=0.85cm}}
		\node[initial,state] (one) {\makecell[c]{$\locone$\\$\mathit{x} \leq 5$\\$\dot{\mathit{x}} \in [a_1,b_1]$}} ; 
		\node[state]		 (two) [right of=one]	{\makecell[c]{$\loctwo$\\$\mathit{x} \leq 10$\\$\dot{\mathit{x}} \in [a_2,b_2]$}}; 
		\path[->]			(one)	edge[bend left] node{\makecell[c]{$\mathit{x} \geq 2.5$}} (two); 
		\path[->]			(two)	edge[bend left] node{\makecell[c]{$\mathit{x} \geq 10$\\$\mathit{x} := 0$}} (one); 
	\end{tikzpicture}%
	\end{adjustbox}%
	%\vspace{-1em}%
	\caption{The hybrid automaton $\AutomatonH$ for~\egref{example}.}
	\figlabel{example_hybrid_automaton}%
	%\vspace{-2em}%
\end{figure}
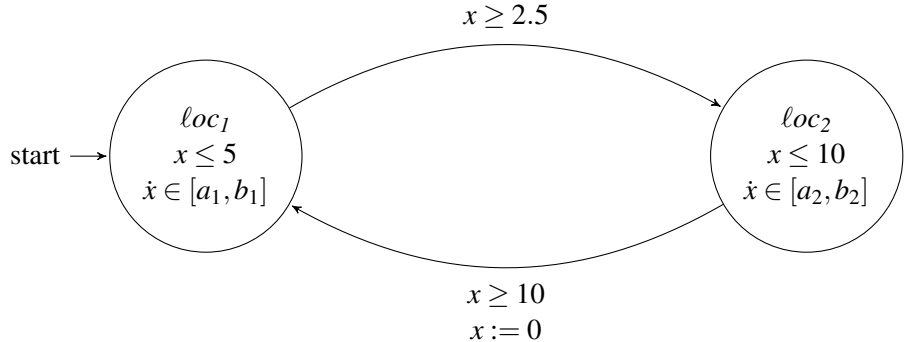%
\eglabel{example}%

%\begin{example}
%
\vspace{1em}
\noindent
{\bf Illustrative Example.}
Consider the hybrid automaton $\AutomatonH$ shown in~\figref{example_hybrid_automaton}.
Assume that the automaton starts at location $\locone$, and the initial value of $\mathit{x}$ is $0$. 
%
%The set of bad states are defined by: $\badset \deq \ \bigvee_{i = 0}^{k} \neg (\qi.\ell_i = \loctwo \ \implies \ \mathit{x} \geq 2.5)$.
%
%
%
Two intervals $[a_1, b_1]$ and $[a_2, b_2]$ describe the rectangular differential inclusions for locations $\locone$, and $\loctwo$, respectively.
This automaton would be a \emph{timed automaton} if all of the constants values are equal, \ie $a_1 = b_1 = a_2 = b_2$.
This automaton would be a \emph{multi-rate timed automaton} if $a_1 = b_1$ and $a_2 = b_2$ but possibly $a_1 \neq a_2$.
Otherwise, this automaton is a \emph{rectangular hybrid automaton}.
Suppose that $a_1 = 1$, $b_1 = 2$, $a_2 = 3$, and $b_2 = 4$.
We introduce $k + 1$ copies $x_0, x_1,...,x_k$ and $\ell_0, \ell_1,...,\ell_k$, where the variable $x_i$ gives the value of the variable $x$, and $\ell_i$ indicates the location at the state $\qi$, representing the $i^{th}$ step of the BMC computation for the automaton shown in~\figref{example_hybrid_automaton}. The set of bad states are defined by: %$\badset \deq \ \bigvee_{i = 0}^{k} \qi.\ell_i = \loctwo \ \wedge \ \mathit{x} < 2.5$.
%\begin{align*}
%\badset \deq \ \bigvee_{i = 0}^{k} \qi.\ell_i = \loctwo \ \wedge \ \mathit{x} < 2.5. 
%\end{align*}

\begin{equation} 
\badset \deq \ \bigvee_{i = 0}^{k} \qi.\ell_i = \loctwo \ \wedge \ \mathit{x} < 2.5. 
\eqlabel{badset_ex}
\end{equation}

The BMC computation of $\AutomatonH$ for each $k$ up to $2$ can be encoded as:
%
%\commenttaylor{I will help fix the next part, it's not correct for the timed part. We have to integrate the ODE to encode the BMC problem and use an existentially quantified amount of real-time to elapse, e.g., suppose $\dot{x}(t) = 1$, then $x(t) = t + x_0$ for some initial value $x_0$. So, to represent the post-states at state $x'$ in the BMC computation, this would be encoded as: $\exists \delta \geq 0$ such that $x' = x + \delta$. To handle the invariant, we would typically use a universal quantifier and require: $\forall \tau 0 \leq \tau \leq t$, $x \leq 5 \wedge x' \leq 5 \wedge x + \delta \leq 5$, which encodes that, for every time $\tau$, the trajectory at time $\tau$, $x(\tau)$, satisfies the invariant, in this case $x(\tau) \leq 5$. Now, for the case of timed and rectangular hybrid automata, it is sufficient to remove this universal quantifier by a convexity argument, which will make the overall query be in a nicer logic (as otherwise there are some $\exists \forall$ quantifier alternations that can get ugly).}
%
\begin{itemize}
	\item $k = 0$: $\iniset_0$ := $(\ell_0 = \locone \wedge \mathit{x}_0 = 0)$;
	%
	% discrete part
	\item $k = 1$ ($\DisTrans_0$): $(\ell_0 = \locone \wedge \ell_1 = \loctwo \wedge x_0 \leq 5 \wedge x_0 \geq 2.5  \wedge x_1 = x_0)$,
	% continuous part
  \item $k = 1$ ($\Trajectory_0$): $(\ell_0 = \locone \implies (\ell_1 = \ell_0 \wedge x_0 + a_1 \delta \leq x_1 \wedge x_1 \leq x_0 + b_1 \delta \wedge x_1 \leq 5))$,
  %
	% discrete part
	\item $k = 2$ ($\DisTrans_1$): $(\ell_1 = \locone \wedge \ell_2 = \loctwo \wedge x_1 \leq 5 \wedge x_1 \geq 2.5  \wedge x_2 = x_1)$,
	% continuous part
  \item $k = 2$ ($\Trajectory_1$): $(\ell_1 = \locone \implies (\ell_2 = \ell_1 \wedge x_1 + a_1 \delta \leq x_2 \wedge x_2 \leq x_1 + b_1 \delta \wedge x_2 \leq 5))$,
	%
%
%\item $T_0$:= $(\ell_0 = \locone \rightarrow \mathit{x}_1 = 0) \wedge (\ell_1 = \locone \rightarrow (\mathit{x}_1 \leq 5) \wedge (a_1 \leq \dot{\mathit{x}_1} \leq b_1))$;
	%
%\item $T(\q_1,\q_2)$:= $(\ell_1 = \locone \rightarrow (\mathit{x}_1 \geq 2.5) \wedge (\mathit{x}_2 = \mathit{x}_1)) \wedge (\ell_2 = \loctwo \rightarrow (\mathit{x}_2 \leq 10) \wedge (a_2 \leq \dot{\mathit{x}_2} \leq b_2))$;
	%
%\item $T(\q_2,\q_3)$:= $(\ell_2 = \loctwo \rightarrow (\mathit{x}_2 \geq 10) \wedge (\mathit{x}_3 = 0)) \wedge (\ell_3 = \locone \rightarrow (\mathit{x}_3 \leq 5) \wedge (a_1 \leq \dot{\mathit{x}_3} \leq b_1))$;
	%	
	%\begin{itemize}
		%\item $\Invset$:= $(\ell = \ell_1 \rightarrow \mathit{x} \leq 5) \wedge (\ell = \ell_2 \rightarrow \mathit{x} \leq 10)$; 
		%
		%\item $\Flowset$:= $(\ell = \ell_1 \rightarrow (a_1 \leq \mathit{x}' \leq b_1)) \wedge (\ell' = \ell_2 \rightarrow (a_2 \leq \mathit{x}' \leq b_2))$;
		%
		%\item $\Transset$:= $(\ell = \ell_1 \rightarrow (\mathit{x} \geq 2.5$ $\wedge$ $\ell' = \ell_2 \wedge \mathit{x}' = \mathit{x})) \wedge (\ell = \ell_2 \rightarrow (\mathit{x} \geq 10 \wedge \ell' = \ell_1 \wedge \mathit{x}' = 0))$.
		%
	%\end{itemize}
\end{itemize}
%
%where $\delta$ is a fresh, real constant\footnote{In general, a universally quantified assertion that the invariant is satisfied for every real time along the trajectory from time $t_0$ to time $t_0 + \delta$, although this is unnecessary for rectangular differential inclusions with linear guards and invariants for convexity reasons~\cite{johnson2012forte,cimatti2015tacas}, which makes this assertion fall into the combination theory of linear real arithmetic with bit-vectors (or some finite sort to encode the locations).}.
where $\delta$ is a fresh, real constant. In general, a universally quantified assertion that the invariant is satisfied for every real time along the trajectory from time $t_0$ to time $t_0 + \delta$, although this is unnecessary for rectangular differential inclusions with linear guards and invariants for convexity reasons~\cite{johnson2012forte,cimatti2015tacas}, which makes this assertion fall into the combination theory of linear real arithmetic with bit-vectors (or some finite sort to encode the locations).
We split the discrete transitions and trajectories for clarity, but the entire formula to be checked for iteration $k = 1$ would just be the disjunction of these conjuncted with the formula representing $k = 0$ and the bad set of states, i.e., $\iniset_0 \wedge (\DisTrans_0 \vee \Trajectory_0) \wedge \badset$.
For $k = 2$, this full formula would be $\iniset_0 \wedge (\DisTrans_0 \vee \Trajectory_0) \wedge (\DisTrans_1 \vee \Trajectory_1) \wedge \badset$.

For $k=1$, we dropped the obviously infeasible transition from $\loctwo$ to $\locone$ from $\DisTrans_0$, which would be found as being unsatisfiable since $\ell_0 \neq \loctwo$.
However, the transition from $\locone$ to $\loctwo$ also cannot occur since $x_0 = 0$, but $x_0 \not\geq 2.5$, so that this part is unsatisfiable and no discrete transitions may be taken from the set of initial states.
We also dropped the continuous dynamics for $\loctwo$ from $\Trajectory_0$ since this would also be infeasible since $\ell_0 \neq \loctwo$.
However, real time may elapse, and as encoded, would correspond to any choice of time $\delta$ such that $x_1 \in [a_1 \delta, b_1 \delta]$ and $x_1 \leq 5$.
Since $a_1 = 1$ and $b_1 = 2$, at most between $2.5$ and $5$ seconds of real time could elapse, and either case would yield $x_1 \in [0, 5]$.

For $k = 2$, we also dropped the infeasible transition and trajectory for clarity.
In this case, the transition from $\locone$ to $\loctwo$ is enabled since $x_1 \in [0, 5]$, so the update to $\loctwo$ may occur.
However, now the continuous trajectory would be infeasible since $x_1$ could already be $5$ and the invariant requires $x_2 \leq 5$, so no real-time $\delta > 0$ may elapse, as otherwise $x_1 + a_1 \delta \leq 5$ is unsatisfiable for $x_1 = 5$.
So, the only state update would be to $\loctwo$ owing to the discrete transition.
%
%This progresses for increasing values of $k$.
%\end{example}

%      
%\vspace{-1em}

%% file: algorithm.tex
\section{Quantified BMC for Hybrid Automata}
\seclabel{algorithm}

The idea of the quantified SMT-based BMC for hybrid automata presented in this paper was inspired from the compact QBF encodings for addressing the BMC problem of purely discrete systems~\cite{jussila2007compressing,janota2012solving, mangassarian2010tc}.
%% 
%%
%
%\subsection{Quantified BMC for Hybrid Automata}
%
To construct a quantified formula $\BMCquantifier$ for the BMC of $\AutomatonH$ of length $k$, we introduce a {\em bit-vector} $\vec{t} = \tuple{t_1, t_2,...,t_{\ceil{\log_2 k}}}$ to index each iteration of the BMC. The next-state of each iteration is connected to the current-state of the next iteration using two multiplexers, where the vector $\vec{t}$ functions as common select lines. We do not describe those multiplexers here but refer to its similar presentation in~\cite{mangassarian2010tc}. Hence, depending on the truth assignment given to the vector $\vec{t}$, the single copy of a transition relation $T$ simulates different iteration of the BMC.   
The QBMC formula is then given as:
\begin{align}
% \BMCquantifier = &\ \exists \q_0, \q_1,\dotline,\qk; \forall t; \exists \q, \q' \mid \iniset(\q_0) \wedge T(\q,\q') \ \wedge \nonumber\\
%& \bigwedge_{i = 0}^{k-1}{{\tiplusone \rightarrow [(\q = \qi) \wedge (\q' = \qiplusone)]}} \wedge \bigvee_{i = 0}^{k}{\badset(\qi)}.
%
\BMCquantifier \deq & \ \exists \VarsetShort_0, \VarsetShort_1,\dotline,\VarsetShort_k,\delta \:\forall \vec{t} \:\exists \VarsetShort, \VarsetShort' \mid \iniset(\VarsetShort_0) \wedge \Transrelation(\VarsetShort,\VarsetShort') \ \wedge \nonumber\\
 & \bigwedge_{i = 0}^{k-1}{{t^{k}(i) \rightarrow [(\VarsetShort = \VarsetShort_i) \wedge (\VarsetShort' = \VarsetShort_{i+1})]}} \wedge (\bigvee_{i = 0}^{k}{\badset(\VarsetShort_i)}),
\eqlabel{bmc_quantifier}
\vspace{1em}
\end{align}
where we note that the existential $\delta$ encodes the real time elapse and would appear in the trajectories $\Trajectory$ of the disjunct $\Transrelation = \DisTrans \vee \Trajectory$.

In the prefix of $\BMCquantifier$, the existential variables $\VarsetShort_i$ is understood to {\em dominate} the universal $\vec{t}$ to ensure state contiguity. Intuitively, if there exists a truth assignment to each existential variable $\VarsetShort_i$, then the quantified first order formulas in SMT will be satisfied for all universal variable assignments.
%\cite{mangassarian2010tc}
Here, the current state $\q$ and the next state $\q'$ under the transition relation $T(\VarsetShort, \VarsetShort')$ are connected to the current state and the next state for each particular iteration $t^{k}(i)$, which is associated with a truth assignment given to $\vec{t}$.  
This allows the QBMC of hybrid automaton $\AutomatonH$ to be compactly encoded using only a single copy of a transition relation with adding some state variable equalities at each time. Whereas, the quantifier-free encoding in \eqref{bmc_smt} requires $k$ copies of the transition relation $T$.
This advantage of non-coping transition relation encoding has been proved to significantly reduce the problem size of the QBF encoding in \cite{mangassarian2010tc}. Similarly, the proposed QBMC method also requires less memory usage compared to other quantifier-free SMT-based BMC approaches, that will be demonstrated in details later in \secref{experiment}. 

To illustrate \eqref{bmc_quantifier}, we consider the QBMC of the hybrid automaton of~\egref{example} with $k = 3$ as follows:
\begin{align} 
 \BMCquantifierthree = &\ \exists \VarsetShort_0, \VarsetShort_1, \VarsetShort_2, \VarsetShort_3, \delta \forall t_1, t_2 \exists \VarsetShort, \VarsetShort' \mid \iniset(\VarsetShort_0) \wedge \Transrelation(\VarsetShort,\VarsetShort')\nonumber\\
	& \wedge \ \{\bar{t_1} \rightarrow [(\VarsetShort = \VarsetShort_0) \wedge (\VarsetShort' = \VarsetShort_1)]\}\nonumber\\
	& \wedge \ \{t_1\wedge \bar{t_2} \rightarrow [(\VarsetShort = \VarsetShort_1) \wedge (\VarsetShort' = \VarsetShort_2)]\}\nonumber\\
	& \wedge \ \{t_1 \wedge t_2 \rightarrow [(\VarsetShort = \VarsetShort_2) \ \wedge (\VarsetShort' = \VarsetShort_3)]\}\nonumber\\
	& \wedge \ (\badset(\VarsetShort_0) \ \vee \badset(\VarsetShort_1) \ \vee \badset(\VarsetShort_2) \ \vee \badset(\VarsetShort_3)),
\eqlabel{example_bmc_quantified}
\end{align} 
where $\VarsetShort = \VarsetShort'$ is a shorthand indicating every variable $v \in \VarsetShort$ equals its corresponding counterpart $v' \in \VarsetShort$.
In~\eqref{example_bmc_quantified}, if the value of $t_1$ is $0$, then there is a continuous trajectory that evolves from the initial state $\q_0$, where $\q_0.\ell_0 = \locone$ and $\mathit{x}_0 = 0$, to the next state $\q_1$, where $\q_1.\ell_1 = \locone$ and $\mathit{x}_1 \leq 5$.
When $t_1 = 1$ and $t_2 = 0$, the system takes the discrete transition from the current state $\q_1$ to the next state $\q_2$, where $\q_2.\ell_2 = \loctwo$ and the value of $\mathit{x}_2$ is not higher than 10. 
At $k = 3$, both $t_1$, and $t_2$ are true, then $\q_2$ becomes the current state, and $\q_3$ is the next state, where $\q_3.\ell_3 = \locone$, and $\mathit{x}_3 \leq 5$. The discrete transition taken from $\q_2$ to $\q_3$ when $\mathit{x} \geq 10$ will reset the value of $\mathit{x}$ to 0.   
%
%
%\begin{itemize}
	%\item $\iniset(\q_0)$:= $\ell = \ell_1 \wedge \mathit{x} = 0$;
	%	
	%\item $\Tone \wedge \Ttwo$:= $\Invset \wedge \Flowset \wedge \Transset $, where:
	%	
	%\begin{itemize}
		%\item $\Invset$:= $(\ell = \ell_1 \rightarrow \mathit{x} \leq 5) \wedge (\ell = \ell_2 \rightarrow \mathit{x} \leq 10)$; 
		%
		%\item $\Flowset$:= $(\ell = \ell_1 \rightarrow (a_1 \leq \mathit{x}' \leq b_1)) \wedge (\ell' = \ell_2 \rightarrow (a_2 \leq \mathit{x}' \leq b_2))$;
		%
		%\item $\Transset$:= $(\ell = \ell_1 \rightarrow (\mathit{x} \geq 2.5$ $\wedge$ $\ell' = \ell_2 \wedge \mathit{x}' = \mathit{x})) \wedge (\ell = \ell_2 \rightarrow (\mathit{x} \geq 10 \wedge \ell' = \ell_1 \wedge \mathit{x}' = 0))$.
		%
	%\end{itemize}
%\end{itemize}
%\commentluan{adding the quantified BMC of $\AutomatonH$ with $k = 3$ in~\egref{example}}.

%
% formula $\BMC$ in~\eqref{example_bmc_quantified} is encoded using 
%linear real arithmetic (LRA), nonlinear real arithmetic (NRA), and fixed size bit-vectors (BV).

If it terminates, an SMT solver supporting the combined theory of bitvectors and reals with quantifiers will return SAT for the QBMC formula if there exists an execution from an initial state to a bad state, \ie if a bad state is reachable.
Otherwise, if it terminates, it will return UNSAT if a bad state is not reachable in $k$ steps.
We note that the combination theory of linear real arithmetic with bitvectors is decidable, and Z3 is in essence a decision procedure for this theory.
%
%, and Z3 also has a decision procedure for nonlinear real arithmetic.
%
%
For instance, the example from~\eqref{example_bmc_quantified} has the state $\q$, where $\q.\ell = \loctwo$, and $\mathit{x} < 2.5$ is unreachable. 
Up to $k = 3$, if the system has any safety specification containing the state $\q$, that means $\q \models \bigvee_{i = 0}^{3}{\badset(\qi)}$, and an SMT solver will return SAT on checking the formula $\BMCquantifierthree$.
Thus, we can assert that the system is unsafe for $k = 3$, which in this case, was just illustrating a reachability query.
%

%\commenttaylor{Some of the previous is vague. We need to be very clear that the quantifiers are used and where, as this is the whole point of this paper. This paper is not about general ``SMT'' encoding of verification problems for hybrid automata, this has already been done (e.g., in my thesis and many other places), it is about using a quantified encoding of BMC, which under the hood, relies on querying SMT solvers. An SMT solver does not have to support quantifiers, but some do, such as Z3.}
%

%% file: experiment.tex
\section{Experimental Results}
\seclabel{experiment}
\newcommand{\timeout}[0]{T/O\xspace}
\newcommand{\memout}[0]{M/O\xspace}
\newcommand{\nomem}[0]{N/A\xspace}
\newcolumntype{C}[1]{>{\centering\let\newline\\\arraybackslash\hspace{0pt}}m{#1}}
%
%In this section, we demonstrate the applicability of the proposed QBMC approach on s
We implement the QBMC method described in \secref{algorithm} as a module within HyST~\cite{bak2015hscc}.
HyST takes as input a hybrid automaton model in an extended form of the SpaceEx XML format~\cite{frehse2011cav} (supporting \eg nonlinear functions instead of only affine ones). Then it will generate a Python script that includes the transition relations of a hybrid system expressed as quantified SMT formulas using the Z3 Python API. %Then one can run the script with Python.
%
%\footnote{The preliminary implementation described in this paper, along with all the examples, is available online at: \url{https://github.com/LuanVietNguyen/QBMC}.}.
%
We evaluate the proposed QBMC method on several instances of Fischer and Lynch-Shavit mutual exclusion protocols. We compare the results to that of dReach, which is a state-of-the-art BMC tool for nonlinear hybrid automata~\cite{gao2013cade}, and with that of HyComp using the MathSAT SMT solver~\cite{cimatti2015tacas} in terms of execution time and memory consumption. 
We note that the comparison focuses only the BMC feature of dReach and HyComp, and all of the models of them were generated using HyST.  
The experiments are performed on Intel I5 2.4GHz processor with 4GB RAM, executing the method described in this paper and dReach in a VirtualBox virtual machine running Ubuntu 64-bit.
Z3 version 4.3.2 was used in the evaluation.
We collect the execution times in second and the peak memory usages in MB for different examples. The preliminary implementation described in this paper, along with all the examples, is available online at: \url{https://github.com/LuanVietNguyen/QBMC}

\begin{table}[t!]
\setlength{\tabcolsep}{1.2em} % for the horizontal padding
\renewcommand{\arraystretch}{2}% for the vertical padding
%\scriptsize
%\vspace{-1em}
\centering
%\begin{adjustbox}{max size={0.95\columnwidth}{0.75\textheight}}%
\caption{The performance comparison between QBMC, HyComp and dReach in solving the BMC of \egref{example}. $k$ is a number of steps in the BMC computation. ``Time" and ``Mem" are the time and memory usages measured in second and MB, respectively.\vspace{2em}
}
%\begin{tabular}{|c|c|C{0.7cm}|C{0.7cm}|C{0.7cm}|C{0.7cm}|C{0.7cm}|C{0.7cm}|}
\begin{tabular}{|c|c|c|c|c|c|c|}
	\hline
	\multirow{2}{*}{Tools} &
	\multicolumn{2}{c|}{k $\leq$ 32} &
	\multicolumn{2}{c|}{k $\leq$ 64} &
	\multicolumn{2}{c|}{k $\leq$ 128} \\
	\cline{2-7}
	& Time & Mem & Time &  Mem & Time &  Mem \\
	\hline
	%\toolQBMC & 2 & 1.11 & 27.2 & 3.68 & 39.4 & 19.9 & 91.2\\
	\toolQBMC & 0.81 & 17.3 & 3.22 & 27.7 & 5.36 & 41.2\\
	\hline
	%dReach  & 2 & 86.7 & 102.4 & 1176.4 & 284.7 & 20034 & 829.2 \\
	dReach  & 82.6 & 106.7 & 1128.2 & 275.5 & 19896 & 812.6 \\
	\hline
	%HyComp  & 2 & 0.4 & 97.3 & 0.6 & 101.8 & 1.44 & 109.3\\
	HyComp  & 0.4 & 98.1 & 0.6 & 105.8 & 1.62 & 128.7\\
	\hline
\end{tabular}
%\end{minipage}
\tablabel{examples_table}
%\vspace{2em}
%\end{adjustbox}
\end{table}   

We first evaluate our QBMC encoding on the illustrative hybrid automata presented in~\egref{example}, and compare the results to those of dReach and HyComp.
%
%, where \toolQBMC denotes the QBMC presented in this paper, $k$ is a number of steps in the BMC computation, and $L$ is the number of discrete locations.
%
The performances of those three different methods are shown in \tabref{examples_table}.
The constant values of the rectangular differential inclusion dynamics are given as: $a_1 = 0, b_1 = 1, a_2 = 0$, and $b_2 = 2$.
The set of bad state used in this experiment is given in \eqref{badset_ex}, where all states in this set are unreachable and the system is always safe. 
%
%: $\phi \deq \ \forall i \in \{1, \ldots, k\}\mid \qi.\ell_i = \loctwo \ \rightarrow \ \mathit{x} < 2.5$, then all states in this set are unreachable and the system is always safe. 
%
For all of the values of $k$ up to 128, the BMC results of QBMC, HyComp and dReach are all UNSAT, illustrating the correctness of the BMC procedure.
The time and memory consumptions shown in~\tabref{examples_table} preliminarily indicate that our QBMC approach is capable of solving BMC faster than dReach, but slower than HyComp.
However, our approach is more scalable as it requires significantly less memory usage compared to dReach and HyComp.

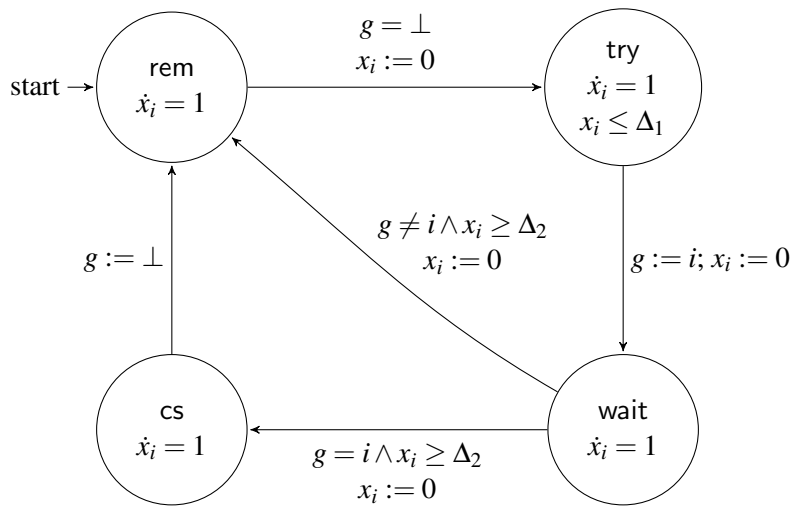
\begin{figure}[t!]
	\vspace{2em}
	\centering
	\scriptsize
	%\begin{adjustbox}{max size={0.95\columnwidth}{0.75\textheight}}
	%\begin{tikzpicture}[>=stealth',shorten >=1pt,auto,node distance=3.5cm]
	    %\tikzstyle{every state}=[rectangle,rounded corners]
	\begin{tikzpicture}[-,>=stealth',shorten >=1pt,auto,node distance=6cm,main node/.style={circle,fill=black!20,draw}]
		\tikzset{every state/.style={minimum size=2cm}, font=\normalsize}
			%\tikzstyle{every state}=[circle]
		\node[initial,state] (rem)      			{\makecell[c]{$\locrem$\\$\localvardot{x}{i} = 1$}};
		\node[state]		 (try) [right of=rem]	{\makecell[c]{$\loctry$\\$\localvardot{x}{i} = 1$\\$\localvar{x}{i} \leq \Delta_1$}};
		\node[state]		 (wait) [below=25mm of try]	{\makecell[c]{$\locwait$\\$\localvardot{x}{i} = 1$}};
		\node[state]		 (cs) at (rem |- wait)	
{\makecell[c]{$\loccs$\\$\localvardot{x}{i} = 1$}};
		\path[->]			(rem)	edge   node {\makecell[c]{$g = \bot$\\$\localvar{x}{i} := 0$}} (try);
		\path[->]			(try)	edge   node {$g := i$; $\localvar{x}{i} := 0$} (wait);
		%\path[->]			(wait)	edge   node {$g \neq i \wedge \localvar{x}{i} 
%\geq B$\\$\localvar{x}{i} := 0$} (rem);
		\path[->]			(wait)	edge[out=150,in=320]   node [above,xshift=10mm,yshift= -2mm] {\makecell[c]{$g \neq i \wedge \localvar{x}{i} \geq \Delta_2$\\$\localvar{x}{i} := 0$}} (rem);
		\path[->]			(wait)	edge   node {\makecell[c]{$g = i \wedge \localvar{x}{i} \geq \Delta_2$\\$\localvar{x}{i} := 0$}} (cs);
		\path[->]			(cs)	edge   node {$g := \bot$} (rem);
	\end{tikzpicture}
	%\end{adjustbox}%
	\caption{Fischer's mutual exclusion algorithm for a process with identifier $i \in \{1, \ldots, N\}$. Here, $g$ is a global variable of type $\{\bot, 1, \ldots, N\}$, $\localvar{x}{i}$ is a local variable of type $\mathbb{R}$, and both $\Delta_1$ and $\Delta_2$ are constants of type $\mathbb{R}$.}%
%	\vspace{-1em}%
	\figlabel{code:fischer}%
\end{figure}

\subsection{Fischer mutual exclusion protocol}
%\vspace{1em}
%\noindent {\bf Fischer mutual exclusion protocol.}
Next, we evaluate \toolQBMC with several scenarios using the Fischer mutual exclusion protocol~\cite{johnson2012forte}.
%
%
%
%dutertre2004timed
%
Fischer mutual exclusion is a timed distributed algorithm that ensures a mutual exclusion safety property, namely that at most one process in a network of $N$ processes may enter a critical section simultaneously. \figref{code:fischer} shows the model of Fischer protocol in including four discrete locations $\Loc \deq \{\locrem, \loctry, \locwait, \loccs\}$, where $\Delta_1$ and $\Delta_2$ are two real timing parameters. Here, if $\Delta_1 < \Delta_2$ a mutual exclusion is guaranteed.
The set of bad states is defined by: 
\begin{align*}
\phi \ \deq \ \neg \forall i, j \in \{1, \ldots, N\} \ | \ (i \neq j \wedge \ q_i = cs) \ \implies \ q_j \neq cs, 
\end{align*}
where $q_i$ and $q_j$ are variables modeling the discrete location of the automata, $cs$ is the critical section location. %and $\rightarrow$ is logical implication.
To evaluate \toolQBMC, we perform the BMC for both of the safe and unsafe version of Fischer protocol. In the safe version, a state where the set of bad states $\phi$ is satisfied is not reachable, while in the unsafe one, a state where $\phi$ is satisfied is reachable.
We then also compare the performance of \toolQBMC in solving the BMC of Fischer protocol with HyComp and dReach. 
%The results of runtime and memory usage the BMC of Fischer protocol in HyComp are shown in~\tabref{trials_table_HyComp}.  
% 
%\figref{runtime_memory} shows the runtime and memory usage comparison among HyComp, dReach and \toolQBMC for different numbers of processes of Fischer protocol; where \toolQBMC-safe, \toolQBMC-unsafe, HyComp-safe, HyComp-unsafe, dReach-safe, and dReach-unsafe denote the BMC of the safe and unsafe version of Fischer protocol using \toolQBMC, HyComp, and dReach, respectively. 
%
%
% 
%
%
%
%The details of run times and memory usages of BMC for the Fischer protocol using these tools are shown in~

\begin{sidewaystable}%[t!]
\centering
\caption{The performance of the BMC of Fischer mutual exclusion protocol using \toolQBMC, HyComp, and dReach.}  \vspace{2em}
\setlength{\tabcolsep}{0.8em} % for the horizontal padding
{\renewcommand{\arraystretch}{1.5}}% for the vertical padding
\begin{tabular}{|c|c|c|c|c|c|c|c|c|c|c|c|}
\hline
\multirow{2}{*}{Example} & \multirow{2}{*}{NoL} & \multirow{2}{*}{k ($\leq$) } & \multicolumn{3}{c|}{\toolQBMC} & \multicolumn{3}{c|}{HyComp} & \multicolumn{3}{c|}{dReach} \\
\cline{4-12}
                         &                      &                    											& Time   & Mem   & Result   & Time   & Mem   & Result   & Time   & Mem   & Result   \\
\hline
\multirow{2}{*}{\Fischerunsafetwo}        & \multirow{4}{*}{$4^2$}    & 8                 &  1.1      &   24.7    &    SAT      &  0.4      &   100.9    &    SAT      &   48.4     &   28.9    &  SAT        \\
\cline{3-12}
                         &                      											& 16                &  1.52      &  28.2     &   SAT       &   0.5     &  101.4     &     SAT      &  50.3      &  30.7     & SAT         \\
\cline{1-1}\cline{3-12}
%\hline
\multirow{2}{*}{\Fischersafetwo}        &     & 8                  												&  1.6      &  25.2     &   UNSAT       & 0.5       & 101.4      &   UNSAT       &  64.1      &  120.8    &    UNSAT       \\
\cline{3-12}
                         &                      &16                 											&  6.4      &   30    &      UNSAT      & 2.8       & 107.3      &   UNSAT       &  \timeout      &  \nomem     &  $\times$        \\
\hline
\multirow{2}{*}{\Fischerunsafethree}        & \multirow{4}{*}{$4^3$}    & 8               &  6.9     &   48.7    &   UNSAT         &  2.1      & 131.8     &  UNSAT        &  270      &  214     &   UNSAT       \\
\cline{3-12}
                         &                      & 16                 											&  22.7     &  49.7     &   SAT        &  6.7      &  149.6     &    SAT       &    959.3     &  235.3      &     SAT      \\
\cline{1-1}\cline{3-12}
\multirow{2}{*}{\Fischersafethree}        &     & 8                  											&  8.3     &   48.7    &    UNSAT        &  2.2      & 131.8       &  UNSAT        &   \timeout       &  \nomem     &   $\times$     \\
\cline{3-12}
                         &                      & 16                 											&  52.4      &  52.4     &    UNSAT        &  55.8      &  214.4      & UNSAT         &       \timeout      &  \nomem     &  $\times$     \\
\hline
\multirow{2}{*}{\Fischerunsafefour}        & \multirow{4}{*}{$4^4$}    & 8                &  40.1      &  73.2     &   UNSAT         &  13.3      & 318.2      & UNSAT         &        \timeout      &  \nomem     &  $\times$        \\
\cline{3-12}
                         &                      & 16                 											&  119.1      &  156.2     &   UNSAT         &  569.4      &   895.4    &  UNSAT        &       \timeout      &  \nomem     &  $\times$         \\
\cline{1-1}\cline{3-12}
\multirow{2}{*}{\Fischersafefour}        &     & 8                  											&  76.1      &  74.1     &     UNSAT       &  9.9      &  319.1     &  UNSAT        &   \timeout      &  \nomem     &  $\times$         \\
\cline{3-12}
                         &                      & 16                 											&  \timeout      &  \nomem     &  $\times$        &   788     &  1010.4     &   UNSAT       &   \timeout      &  \nomem     &  $\times$         \\
\hline
\multirow{2}{*}{\Fischerunsafefive}        & \multirow{4}{*}{$4^5$}    & 8                &  288.8      &   249.9    &    UNSAT       &  109.1      &   1345.4    &    UNSAT      &   \timeout      &  \nomem     &  $\times$       \\
\cline{3-12}
                         &                      & 16                 											&  21456     &   473.8    &     UNSAT     &   \nomem      &    \memout    &    $\times$       &  \timeout      &  \nomem     &  $\times$       \\
\cline{1-1}\cline{3-12}
\multirow{2}{*}{\Fischersafefive}        &     & 8                  											&  344.4      &  254.4     &   UNSAT      &  172.4      &    1405.9   &   UNSAT       &    \timeout      &  \nomem     &  $\times$        \\
\cline{3-12}
                         &                      & 16                 											&  \timeout      &   \nomem     &   $\times$   & \nomem      &  \memout      &   $\times$             &    \timeout      &  \nomem     &  $\times$          \\
\hline
\end{tabular}
\tablabel{trials_table_2}
\end{sidewaystable}

\tabref{trials_table_2} shows the execution time and memory usage comparison among HyComp, dReach and \toolQBMC for Fischer protocol with different numbers of processes, where NoL is the number of locations; \Fischersafe, \Fischerunsafe denote the safe and unsafe versions of Fischer protocol, respectively, and the number following the hyphen (-) describes a number of processes for each version.
%
%In \Fischersafe, a state where the set of bad states $\phi$ is satisfied is not reachable, while in \Fischerunsafe, a state where $\phi$ is satisfied is reachable.
%
For instance, \Fischersafetwo, \Fischerunsafetwo are the safe and unsafe versions of the Fischer protocol with $2$ processes, respectively. We choose $\Delta_1 = 5$, $ \Delta_2 = 70$ for a safe version, and $\Delta_1 = 75$, $\Delta_2 = 70$ for an unsafe one.
%
%\tabref{trials_table_2} shows that the BMC of Fischer protocol with 64 discrete locations can be checked completely up to $k = 32$.
%\tabref{trials_table_2}
In \tabref{trials_table_2}, \timeout means the computation time out ($\geq$ 12 hours), \memout represents that the peak memory usage is higher than 4GB, and \nomem denotes that the information of times or memory usages cannot be captured due to either \timeout or \memout. Also, $\times$ means that the BMC procedure cannot terminate.

\paragraph*{Compared to HyComp.} According to \tabref{trials_table_2}, we can see that HyComp is generally faster than \toolQBMC, but it requires a higher memory consumption than \toolQBMC.
For instance, up to $k = 8$, HyComp can address the BMC of the safe version of Fischer protocol with 5 processes in 172.4 seconds, which is approximately twice faster than \toolQBMC (terminates after 344 seconds). However, HyComp consumes 1405.9 MB of memory, which is more than five times higher than QBMC (using only 254.4 MB). 
Additionally, with $k \leq 16$, the BMC of the unsafe version of Fischer protocol with 5 processes cannot terminate in HyComp due to out of memory (requiring more than 4GB).
% HyComp can address the BMC of the safe version of Fischer protocol with 5 processes in 172.4 seconds, which is approximately twice faster than \toolQBMC (terminates after 344 seconds)
However, \toolQBMC can solve it using less than 500 MB. %($> 85\%$ memory reduction).
Hence, we can deduce that \toolQBMC is superior than HyComp with respect to memory consumption.
%
%Moreover,~\figref{runtime_memory} also indicates that \toolQBMC is able to solve BMC of hybrid automata faster and uses much less memory than dReach.
%

\paragraph*{Compared to dReach.} Generally, \tabref{trials_table_2} also indicates that \toolQBMC outperforms dReach in solving the BMC of Fischer protocol. As an example, with $k \leq 16$,  dReach solves the BMC of the unsafe version of Fischer protocol with 3 processes in 959.3 seconds using 235.3 MB of memory, while \toolQBMC only terminates in 22.7 seconds ($\approx$ 40 time faster) using only 49.7 MB. Moreover, the BMC of both unsafe and safe versions of Fischer protocol with more than 3 processes is not able to terminate in dReach. Although dReach is capable to address the BMC of a wide range of nonlinear hybrid systems, handling scalability is not its strength. 

%($\approx 4000\%$ runtime reduction) ($\approx 500\%$ memory reduction)
Overall, \toolQBMC has a competitive execution time and is more scalable to other state-of-the-art SMT solvers.
Due the state-space (and formula) explosion of BMC, the reduction of memory consumption is one of the major challenges to address.
Since \toolQBMC requires a smaller amount of memory usage than other quantifier-free BMC approaches, it is effective in solving the BMC of large scale problems. 
According to \tabref{trials_table_2}, the BMC of the unsafe version of 4-processes Fischer protocol can be checked completely using QBMC up to $k = 32$ with only 254.1 MB of memory consumption. This result indicates that \toolQBMC is effective for bug detection.
However, as $k$ increases, the higher execution time and the larger memory usage are required for the quantified encoding of BMC due to the increasing number of all possible paths from an initial state in the set of initial states to a bad state that does not satisfy the set of safety specifications.

\subsection{Lynch-Shavit mutual exclusion protocol}
%\vspace{1em}
%\noindent {\bf Lynch-Shavit mutual exclusion protocol.}
%We also evaluate \toolQBMC with the Lynch-Shavit mutual exclusion protocol.
%
The Lynch-Shavit protocol is a modified version of Fischer protocol where the mutual exclusion property is time-independent\cite{lynch1992timing}. Intuitively, the protocol ensures that mutual exclusion is always satisfied even if the timing constraints are violated.
%~\cite{alur1992results}.
%
We first modeled the hybrid automata of Lynch-Shavit protocol~\cite{abdulla2004using} in an extended form of the SpaceEx XML format. Then, we used the HyST tool to translate it to other model formats (e.g., for HyComp and dReach), and create the QBMC encoding.
Each process of Lynch-Shavit protocol has 9 discrete locations, so the Lynch-Shavit protocol with 4 processes includes 6561 locations.
%
%It is a challenge for SMT solvers to address the BMC of a system included that such large number of states. 
%
%\memout presents that the peak memory usage is higher than 3GB, and \nomem denotes that the information of runtimes is not detected due to \memout.
%
The set of bad states of Lynch-Shavit protocol is defined similar to Fischer protocol, where two processes may be in the critical section.

The performance analyzing the BMC of Lynch-Shavit protocol using \toolQBMC and HyComp are shown in~\tabref{lynch_table}, in which the BMC results of QBMC and HyComp are both UNSAT. Due to scalability problem, the BMC of Lynch-Shavit protocol using dReach could not terminate correctly. Thus, we do not demonstrate the time and memory consumptions for the BMC of Lynch-Shavit protocol using dReach in~\tabref{lynch_table}. However, we still provide the model files in our preliminary implementation package for interested readers to investigate. 
Again, we can see the trade-off between the two approaches.
HyComp is generally faster than \toolQBMC, but requires a much higher memory usage.
As a result, the BMC of Lynch-Shavit protocol with 4 processes can be solved by \toolQBMC up to $k = 16$ with consuming approximately 1GB memory, but cannot be solved in HyComp up to $k = 8$ due to out of memory. 
Hence, that demonstrates the proposed QBMC approach is efficient and scalable in addressing the BMC of large hybrid systems.

\begin{table}[t!]
\setlength{\tabcolsep}{1.2em} % for the horizontal padding
\renewcommand{\arraystretch}{2}% for the vertical padding
	%\scriptsize
	\centering
	%\begin{adjustbox}{max size={0.95\columnwidth}{0.75\textheight}}%
	\caption{The performance comparison between \toolQBMC and HyComp in solving the BMC of Lynch-Shavit mutual exclusion protocol. 
	}
	\vspace{2em}
%\begin{tabular}{|c|c|C{0.7cm}|C{0.7cm}|C{0.7cm}|C{0.7cm}|C{0.7cm}|C{0.7cm}|C{0.7cm}|}
	\begin{tabular}{|c|c|c|c|c|c|c|c|c|}
		\hline
		\multirow{2}{*}{Tools} &
		\multirow{2}{*}{NoL} & 
		\multicolumn{2}{c|}{k $\leq$ 4} &
		\multicolumn{2}{c|}{k $\leq$ 8} &
		\multicolumn{2}{c|}{k $\leq$ 16} \\
		\cline{3-8}
		& &Time & Mem & Time &  Mem & Time &  Mem\\
		\hline
		\multirow{3}{*}{\toolQBMC} 
		& $9^2$ & 3.7 & 52.2 & 5.1 & 52.3 & 25.9 & 52.7\\
		\cline{2-8}
		& $9^3$ & 15.5 & 65.6 & 31.3 & 87.5 & 1091.5 & 144.5\\
		\cline{2-8}
		& $9^4$ & 256.1 & 702.8 & 1062.1 & 708.9 & 43178 & 1196.2\\
		\hline
		\multirow{3}{*}{HyComp} 
		& $9^2$ & 0.8 & 121.9 & 1.33 & 132.8 & 9.5 & 170.5\\
		\cline{2-8}
		& $9^3$ & 2.7 & 307.9 & 12.81 & 380.8 & 192.8 & 771.4\\
		\cline{2-8}
		& $9^4$ & 63.9 & 2655.4 & \nomem & \memout & \nomem & \memout\\
		\hline
	\end{tabular}
	%\end{minipage}
	\tablabel{lynch_table}
	%\vspace{-2em}
	%\end{adjustbox}
\end{table}

%% file: conclusion.tex
\section{Conclusion and Future Works}
\seclabel{conclusion}
%\vspace{-0.5em}
%
In this paper, we present a new SMT-based technique that encodes, in a quantified form, the BMC problem for rectangular hybrid systems which also subsumes this encoding for timed systems.
The preliminary results for the Fischer mutual exclusion protocol and Lynch-Shavit protocol indicate the capability of our method to solve the BMC problem for hybrid systems including more than thousand locations.
We compare the QBMC approach to the quantifier-free BMC approaches in the dReach tool that uses the dReal SMT solver, and the HyComp tool built on top of nuXmv that uses the MathSAT SMT solver. The experimental results demonstrate that our approach is competitive to these tools in terms of execution time and memory consumption.
%
%
%
%

%\vspace{1em}
\emph {Future works.} As solvers for fragments of many-sorted first-order logic such as LRA, NRA, etc., continue to improve, QBMC encodings such as the one described in this paper will become more effective, similar to how QBMC for discrete systems has been shown to be effective with QBF encodings~\cite{mangassarian2010tc}. In future, we plan to conduct additional experiments for other benchmarks to cover a wider range of applications, and compare the results to more tools and techniques, such as Uppaal and tools built on top of CVC4~\cite{DBLP:conf/cav/BarrettCDHJKRT11}. We intend to compare the proposed QBMC with IC3-based algorithms that might provide trade-offs between memory consumption and execution time. We will also investigate more general classes of hybrid automata whose dynamics are expressed by linear or polynomial differential equations.